\documentclass[twoside, a4paper, notoc, 10pt ]{JHEP3}
\usepackage{amsfonts}
\usepackage{amssymb}
\usepackage{comment}
\usepackage{epsfig}
\usepackage{graphicx}
\usepackage{amsmath}
\usepackage{amsxtra}
\usepackage{cancel}

\newcommand{\mbb}{\mathbb}
\newcommand{\mc}{\mathcal}

\newcommand{\roughly}[1]{\mathrel{\raise.3ex\hbox{$#1$\kern-0.85em
\lower1ex\hbox{$\sim$}}}}

\newcommand{\lsim}{\roughly<}
\newcommand{\gsim}{\roughly>}

\newcommand{\bea}{\begin{eqnarray}}
\newcommand{\eea}{\end{eqnarray}}
\newcommand{\vo}{{\cal V}}
\newcommand{\be}{\begin{equation}}
\newcommand{\ee}{\end{equation}}

\def\ba{\begin{eqnarray}}
\def\ea{\end{eqnarray}}

\def\ssD{{\scriptscriptstyle D}}

\def\ssF{{\scriptscriptstyle F}}

\def\ssM{{\scriptscriptstyle M}}
\def\ssN{{\scriptscriptstyle N}}

\def\ssW{{\scriptscriptstyle W}}
\def\KK{{\scriptscriptstyle KK}}
\def\SM{{\scriptscriptstyle SM}}
\def\JF{{\scriptscriptstyle J \kern-0.15em F}}
\def\EF{{\scriptscriptstyle E \kern-0.15em F}}

\def\cO{\mathcal{O}}

\def\cL{\mathcal{L}}
\def\cN{\mathcal{N}}

\def\cV{\mathcal{V}}

\def\nn{\nonumber}

\def\({\left(}
\def\){\right)}

\def\exd{{\rm d}}

\def\pref#1{(\ref{#1})}

\title{Anisotropic Modulus Stabilisation: Strings at\\ LHC Scales with
Micron-sized Extra Dimensions}

\author{
M.~ Cicoli,${}^1$ C.P.~Burgess,${}^{2,3}$ F.~Quevedo${}^{4,5}$
\\ $^1$Theory Group, Deutsches Elektronen-Synchrotron DESY, D-22603 Hamburg, Germany \\
$^2$ Department of Physics \& Astronomy, McMaster University,
Hamilton ON, Canada.\\
$^3$ Perimeter Institute for Theoretical Physics,
 Waterloo ON, Canada.\\
$^4$ DAMTP/CMS, University of Cambridge, Cambridge CB3 0WA, UK.\\
$^5$ Abdus Salam ICTP, Strada Costiera 11, Trieste 34014, Italy.}

\abstract{We construct flux-stabilised Type IIB string compactifications whose extra dimensions have very different sizes, and use these to describe several types of vacua with a TeV string scale. Because we can access regimes where two dimensions are hierarchically larger than the other four, we find examples where two dimensions are micron-sized while the other four are at the weak scale in addition to more standard examples with all six extra dimensions equally large.
Besides providing ultraviolet completeness, the phenomenology of these models is richer than vanilla large-dimensional models in several generic ways: ($i$) they are supersymmetric, with supersymmetry broken at sub-eV scales in the bulk but only nonlinearly realised in the Standard Model sector, leading to {\em no} MSSM superpartners for ordinary particles and many more bulk missing-energy channels, as in supersymmetric large extra dimensions (SLED); ($ii$) small cycles in the more complicated extra-dimensional geometry allow some KK states to reside at TeV scales even if all six extra dimensions are nominally much larger; ($iii$) a rich spectrum of string and KK states at TeV scales; and ($iv$) an equally rich spectrum of very light moduli exist having unusually small (but technically natural) masses, with potentially interesting implications for cosmology and astrophysics that nonetheless evade new-force constraints. The hierarchy problem is solved in these models because the extra-dimensional volume is naturally stabilised at exponentially large values: the extra dimensions are Calabi-Yau geometries with a 4D K3 or $T^4$-fibration over a 2D base, with moduli stabilised within the well-established LARGE-Volume scenario. The new technical step is the use of poly-instanton corrections to the superpotential (which, unlike for simpler models, are likely to be present on K3 or $T^4$-fibered Calabi-Yau compactifications) to obtain a large hierarchy between the sizes of different dimensions. For several scenarios we identify the low-energy spectrum and briefly discuss some of their astrophysical, cosmological and phenomenological implications.\\

{$e$-mail: \email{michele.cicoli@desy.de}; \email{cburgess@perimeterinstitute.ca}; \\ \email{F.Quevedo@damtp.cam.ac.uk} }}

\preprint{DESY 11-073}
\begin{document}

\tableofcontents

\bigskip

\section{Introduction}

The observation that quantum gravity could become important at energies as low as the TeV scale \cite{ADD,ADDpheno} considerably raises the stakes for what might be found at the LHC. Besides its implications for the LHC, if gravity is really a TeV effect it could also imply a variety of novel new non-accelerator phenomena, including modifications to gravity over micron and macroscopic distances and novel cosmology and astrophysics. It raises the prospect of forging a link between astrophysical observations, terrestrial tests of gravity, and collider experiments at very high energies.

In these scenarios predictions for the LHC tend to be quite robust, in that they do not depend strongly on nitty gritty details like the exact shape of the extra dimensions. By contrast, gravitational predictions are much more model-dependent, since they typically probe only the existence and properties of very low-energy states in the sub-eV regime. For instance, in a ten-dimensional world with a TeV gravity scale, predictions for the LHC depend relatively weakly on whether all six dimensions are large or whether two are much larger than the other four. By contrast, observable deviations from gravity on micron scales depend on this very much, since only in the latter case can any dimensions be big enough to be detected. Because of this any real connection between gravity at the LHC and lower-energy observables requires a fairly detailed understanding of the extra dimensions and how they are stabilised.

String theory provides a natural framework for such an understanding, yet detailed mechanisms for stabilising moduli in string theory have been understood in a controlled way only fairly recently, amongst Type IIB Calabi-Yau flux compactifications \cite{GKP,KKLT}. Most interestingly, solutions arise within this framework with the volume, $V_6$, of the extra dimensions naturally stabilised at exponentially large values, $V_6 \propto e^{c/g_s}$, where $g_s \ll 1$ is the string coupling and $c$ is a positive constant of order unity. In particular, relatively small changes to the input parameters can generate the extremely large values,\footnote{We use the reduced Planck scale throughout: $M_p^2 = (8 \pi G)^{-1}$, and so $M_p = 2.4 \cdot 10^{18}$ GeV.}
\be
 \cV := V_6 M_s^6 \propto \frac{M_p^2}{M_s^2} \simeq 10^{30}\,,
\ee
that are required if the string scale, $M_s^{-1} := l_{s} := 2\pi \sqrt{\alpha'}$, is to be as low as: $M_s \sim 1 \,\hbox{TeV}$. These models, called the LARGE volume scenario \cite{LVS,LVSgeneral} or LVS for short, tend to be very predictive, in particular making specific predictions for a rich spectrum of light moduli with masses below the (already small) Kaluza-Klein (KK) scale.

In the simple `Swiss cheese' geometries first studied, TeV strings within the LVS tended to predict similar sizes for all of the extra dimensions, making them all equally large --- {\em i.e.} $L \simeq V_6^{1/6} \sim (10 \, \hbox{MeV})^{-1} \simeq 10$ fm ---  but not so large as the sub-millimetre scales to which tests of Newton's laws are presently sensitive. What is missing so far are models whose extra dimensions are extremely asymmetric in size, with a volume of the form
$V_6 = L^2\,l^4$, where $L \sim 10 \, \mu\hbox{m}\sim (0.01\, \hbox{eV})^{-1} $ is the size of the two large dimensions, and $l \simeq (V_6/L^2)^{1/4} \sim 10^{-4} \, \hbox{fm} \sim (1 \, \hbox{TeV})^{-1} \ll L$ is the size of the other four small dimensions. It is the purpose of this paper to begin filling in this regime, seeking in particular Type IIB models where two dimensions are much larger than the other four.

Having the string scale near a TeV (regardless of whether $L$ differs much from $l$) has crucial implications for how supersymmetry breaks. Although the fluxes in LARGE-volume vacua already break supersymmetry, they do so with a very low scale, $m_{3/2} \simeq M_s^2 /M_p \simeq M_p / \vo$ ($\sim 10^{-3}$ eV when $M_s \sim 1$ TeV). This means that other sources of breaking must dominate in the sector containing standard-model (SM) particles. Since we know this sector must in any case reside on a brane (to prevent having already detected the large dimensions \cite{ADD}), this means that this SM brane must badly break supersymmetry. (Such supersymmetry breaking is quite possible in string theory, such as the explicit local non-supersymmetric brane models constructed in ref.~\cite{singular}.)

This kind of supersymmetry-breaking pattern has two robust consequences. First, it implies some supersymmetry survives down to extremely low energies; with supermultiplets in the bulk split by scales of order $m_{3/2} \sim 10^{-3}$ eV. Indeed we find that the physics that stabilises the extra dimensions robustly predicts yet more states at these same energies. The generic picture has a diverse spectrum of unusually light particles, with potentially rich implications for very-low-energy physics \cite{SLED,MSLED,SLEDpheno,HiddenPhotons}.

Second, despite the low supersymmetry-breaking scale in the bulk, the particle spectrum relevant to LHC physics does {\em not} include the usual superpartners of minimal supersymmetric (MSSM) models. No superpartners arise because on a non-supersymmetric brane supersymmetry relates single-particle to multi-particle states (and so takes an electron, say, to an electron plus a gravitino rather than to the MSSM selectron). Searches for MSSM superpartners at the LHC should come up empty-handed, as indeed they have so far been doing \cite{LHCSUSY}.

In order to find anisotropic stabilisations, in \S2 we explore compactifications that are topologically K3 or $T^4$ fibrations over a $\mathbb{P}^1$ base. We find the moduli of such spaces can stabilise at sufficiently anisotropic shapes to allow the size, $L$, of the base to be of sub-millimetre size. Thus the low-energy limit is described by a 6-dimensional effective field theory (EFT), comprising a stringy derivation of the supersymmetric large-volume scenario \cite{ADD,SLED}. The crucial ingredient for obtaining this is the use of poly-instanton corrections to the superpotential \cite{POLY}. These are instanton-like corrections to the gauge kinetic functions, that contribute to stabilisation through the influence of these kinetic terms on the superpotential. Although usually neglected for modulus stabilisation due to their exponentially small dependence on moduli, they can dominate when the zero-mode structure of a non-rigid K3 or $T^4$ surface forbids single-instanton contributions to the superpotential.

We begin a preliminary exploration of some phenomenological consequences in \S3 and \S4, assuming that the Standard Model itself is localised on $D7$-branes wrapping small cycles within the large overall extra-dimensional volume. We find in \S3 a generic prediction of a rich spectrum of states whose masses are light enough to be relevant to terrestrial tests of gravity, yet which are not already explicitly ruled out. We argue that for very anisotropic compactifications the low-energy world transverse to the branes is effectively 2-dimensional, implying that brane back-reaction is an important complication to the LARGE-volume dynamics at very low energies \cite{6DBraneBR}. On one hand, this puts detailed calculations of the low-energy properties beyond the present state of the art, motivating more detailed a better understanding of back-reaction in hopes of making more precise comparisons with observations. On the other hand, the presence of back-reacting codimension-2 branes might yet be a good thing, since they may provide a new mechanism for understanding the small size of the present vacuum energy \cite{SLED,SLEDrev}.

\S4 provides a preliminary discussion of some of the phenomenological implications. This includes the several ways these string compactifications differ from more naive extra-dimensional phenomenology, as well as distinctive implications for macroscopic tests of gravity. We focus on distinguishing those features that are generic to large-volume and sub-millimetre extra-dimensional models from those more specific to the stabilisation mechanism considered here. Our summary of the results appears in \S5.

\section{Fibred constructions}

This section lays out the guts of our construction. It starts by describing fibred Calabi-Yau geometries, and what an anisotropic compactification looks like when expressed in terms of their moduli. After a brief summary of modulus stabilisation for these geometries, two types of anisotropic stabilisations are described; one relying on string-loop generated interactions, and one relying on poly-instanton interactions. These models differ in the degree of anisotropy obtainable using ordinary input parameters, with the poly-instanton proposal allowing the extreme hierarchies of scale required for micron-sized extra dimensions.

\subsection{Type IIB compactified on fibered Calabi-Yau three-folds} \label{2}

We focus throughout on Calabi-Yau three-folds whose volume can be written in the form
\begin{equation}
 \vo = \lambda_1 t_{1}
 t_{2}^{2} + \lambda_2 t_{3}^{3},
\label{Vol}
\end{equation}
where the $t_i$ are volumes of internal 2-cycles in the geometry, and $\lambda_{1,2}$ are the intersection numbers for these cycles (that depend on the details of the Calabi-Yau of interest). For explicit Calabi-Yau constructions via toric geometry which exhibit this form of the
overall volume see \cite{CKM}.

The volumes, $\tau_i$, of the 4-cycles dual to these 2-cycles are defined by $\tau_i = {\partial \mathcal{V}}/{\partial t_i}$, and so
\begin{equation}
 \tau_1 = \lambda_1 t_2^2,\,\,\,
 \tau_2 = 2\lambda_1t_1t_2,\,\,\,
 \tau_3 = 3\lambda_2 t_3^2 \,.
 \label{taus}
\end{equation}
These define the real part of the geometry's complex K\"{a}hler moduli
\begin{equation}
 T_i = \tau_i + i \int_{D_i} C_4  \,, \,\,
 i=1,...,h_{1,1} = 3 \,,
\label{Tmoduli}
\end{equation}
where $D_i$ is the 4-cycle (divisor) whose volume is given by $\tau_i$, $C_4$ is the
Ramond-Ramond 4-form, and $h_{m,n}$ (with $m,n = 1,2,3$) are the manifold's Hodge numbers. In terms of the $T$-moduli, the volume (\ref{Vol}) reads:
\begin{equation}
 \vo = \alpha \left( \sqrt{\tau_1}\tau_2 - \gamma \tau_3^{3/2}\right)
 = t_1\tau_1-\alpha\gamma\tau_3^{3/2},  \label{hhh}
\end{equation}
where $\alpha $ and $\gamma $ are given in terms of the $\lambda_i$ by: $\alpha = 1/ (2\sqrt{\lambda_1})$ and $\gamma = \frac23 \sqrt{\lambda_1 /(3\lambda_2)}$.

Topologically, this Calabi-Yau three-fold has a $\mbb{P}^1$ base of size $t_1 := \left(L M_s\right)^2$,
a K3 or $T^4$ fibre \footnote{The topology of the fibre can be determined by computing its Euler characteristic $\chi$:
if $\chi=24$ the fibre is a K3 surface whereas if $\chi=0$ the fibre is a $T^4$ surface \cite{Schulz}.} of size $\tau_1 := \left(l M_s\right)^4$ and a point-like singularity resolved
by a blow-up mode whose volume is given by $\tau_3 := \left(d M_s\right)^4$.
For LARGE-volume models we restrict attention to orientifold projections
that project out none of these K\"{a}hler moduli and focus on the large-volume regime, for which
\begin{equation} \label{V0hier}
 t_1 \tau_1 \gg \alpha \gamma \tau_3^{3/2}
 \quad \hbox{in which case} \quad
 \vo \simeq t_1 \tau_1 = L^2l^4 M_s^6\,.
\end{equation}

We seek anisotropic compactifications for which the two dimensions of the base ---
spanned by the cycle $t_1$ --- are hierarchically larger than the four dimensions
of the fibre --- spanned by $\tau_1$, making the base 2-cycle much bigger than its dual 4-cycle. The following sections describe two particular constructions, for which the potential energies are minimised by
\bea
 \hbox{\em Small Hierarchy:} &&\langle t_1 \rangle > \sqrt{\langle\tau_1\rangle} \gg \sqrt{\langle\tau_3\rangle} \quad \hbox{and so} \quad L \gsim l\gg d \,;\nn\\
 \hbox{\em Large Hierarchy:} &&\langle t_1 \rangle \gg \sqrt{\langle\tau_1\rangle} \simeq \sqrt{\langle\tau_3\rangle} \quad \hbox{and so} \quad L \gg l \gsim d \,.\nn
\eea
In our later applications the first of these gives six dimensions that are all at MeV -- GeV scales; the second gives two micron-sized extra dimensions.

\subsection{Boilerplate K\"{a}hler-modulus stabilisation}
\label{3modK3noLoopCalc}

We work within the now-familiar framework of Type IIB string theory compactified with background fluxes sourced by $D7$- and $D3$-branes \cite{GKP}. The 10D theory is IIB supergravity (with orientifold projections such that $h_{1,1}^{-} = 0$), and so the closed-string moduli that require stabilisation include the axio-dilaton, $S = e^{-\phi } + i C_{0}$ (where $\phi$ is the 10D dilaton and $C_0$ the Ramond-Ramond 0-form); a variety of complex-structure moduli, $U_{\alpha}$ (with $\alpha = 1,..., h_{1,2}^{-}$); and the K\"ahler moduli, $T_i$ (with $i = 1,..., h_{1,1}^{+}$ defined in eq.~(\ref{Tmoduli})). Of these, the $S$ and the $U$-moduli can be stabilised at leading order in $g_s$ and $\alpha'$ if nonzero 3-form fluxes are present in the background geometry \cite{GKP,Diego}. By contrast, the K\"ahler moduli $T_i$ remain unstabilised at leading semiclassical order.

The stabilisation of these remaining K\"ahler moduli is more complicated, since it involves dynamics beyond leading order in $g_s$ and $\alpha'$. If this dynamics involves energies smaller than the Kaluza-Klein scale, it can be described in the low-energy effective 4D theory within which the extra-dimensional moduli appear as scalar fields. This effective theory is an effective $\cN = 1$ 4D supergravity (possibly with soft-breaking terms) if the bulk fluxes do not break supersymmetry too badly.

This 4D effective supergravity is described by a K\"{a}hler potential and superpotential that --- at string tree-level and to lowest order in $\alpha'$ --- take the form \cite{CYKahler}
\begin{equation}
  K_{\rm tree} = K_0 -2 \ln \mathcal{V}
  \quad \hbox{and} \quad
  W_{\rm tree} = W_0 \,,
\label{eqtrees}
\end{equation}
where $W_0 = \int\limits G_3 \wedge \Omega$ and $K_0 = - \ln \left( S + \bar{S} \right)  - \ln \left( -i \int\limits \Omega \wedge \bar{\Omega}\right)$ describe the $S$- and $U$-dependent terms, with $U$ appearing through its appearance in the holomorphic (3,0)-form, $\Omega(U)$. Here $G_3$ is the usual IIB complex 3-form flux. These determine (among other things) the $\mc{N}=1$ F-term scalar potential,
\begin{equation}
 V_\ssF =e^K\left[K^{i\bar{j}}\left(W_i+W K_i\right)\left(\bar{W}_{\bar{j}}+\bar{W}K_{\bar{j}}\right)-3|W|^2\right],
\end{equation}
which vanishes identically (as a function of $T_i$) when evaluated using $K_{\rm tree}$ and $W_{\rm tree}$ after $S$ and $U$ are evaluated at their minima.

The K\"ahler metric produced by this K\"ahler potential simplifies considerably in the large-volume limit, which neglects any terms that are subdominant in inverse powers of the two large moduli, $\tau_1$ and $\tau_2$. The leading contribution to the K\"{a}hler metric and its inverse in this limit is
\begin{equation}
 K_{i\bar\jmath}^{0} = \frac{1}{4\tau_2^2}\left(
 \begin{array}{ccccc}
 \frac{\tau_2^2}{\tau_1^2} && \gamma\left(
 \frac{\tau_3}{\tau_1}\right)^{3/2} && -\frac{3 \gamma
 }{2}\frac{\sqrt{\tau_3}}{\tau_1^{3/2}}\tau_2 \\
 \gamma\left(\frac{\tau_3}{\tau_1}\right)^{3/2} && 2 &&
 -3\gamma \frac{\sqrt{\tau_3}}{\sqrt{\tau_1}} \\
 -\frac{3\gamma}{2}\frac{\sqrt{\tau_3}}{\tau_1^{3/2}}\tau_2 && -3\gamma
 \frac{\sqrt{\tau_3}}{\sqrt{\tau_1}}
 && \frac{3\alpha \gamma}{2}\frac{\tau_2^2}{\mathcal{V}\sqrt{\tau_3}}
 \end{array}
 \right), \label{LaDiretta}
\end{equation}
and
\begin{equation}
 K_{0}^{\bar\imath j}=4\left(
 \begin{array}{ccccc}
 \tau_1^2 && \gamma\sqrt{\tau_1}\tau_3^{3/2} &&
 \tau_1\tau_3 \\
 \gamma\sqrt{\tau_1}\tau_3^{3/2} && \frac12
 \, \tau_2^2 && \tau_2 \tau_3 \\
 \tau_1\tau_3 && \tau_2
 \tau_3 && \frac{2}{3\alpha \gamma }\vo\sqrt{\tau _3}
 \end{array}
 \right) \,.  \label{Kinverse}
\end{equation}

\subsubsection*{Perturbative corrections}

But the juice of K\"ahler modulus stabilisation lies not in the above quantities, but rather in the corrections that cause deviations from them. Potentially the most important of these are perturbative corrections in $\alpha'$ and $g_s$, which non-renormalisation theorems \cite{NRtheorems} imply can appear only in $K$.

The leading $\alpha'$ corrections modify $K$ to \cite{bbhl}
\begin{equation} \label{Kalpha}
 K = K_{tree} + \delta K_{(\alpha')} = -2 \ln \left(
 \vo + \frac{\xi}{2 g_s^{3/2}} \right) \,,
\end{equation}
where $\xi$ is given by $\xi ={(h_{1,2}-h_{1,1})\zeta (3)}/{[2(2\pi )^{3}]}$, with $\zeta(3) \simeq 1.2$.

String loops also correct $K$ and the changes that depend on the K\"ahler moduli typically arise from open-string loops; they depend on the moduli of the cycles on which the corresponding $D7$-branes are wrapped. As a result the precise form of the correction, $\delta K_{(g_s)}$, depends on the details of which branes wrap which cycles. A few examples suffice to indicate the kinds of one-loop contributions that can arise.

\begin{enumerate}
 \item {\em $D7$s wrapping $\tau_1$ and $\tau_2$; $ED3$ wrapping $\tau_3$:} For instance, suppose we wrap a stack of spacetime-filling $D7$-branes -- denoted $D7_1$ -- around the 4-cycle $\tau_1$ in the fibred Calabi-Yau considered above; and wrap another stack -- $D7_2$ -- around the cycle $\tau_2$. Finally suppose a Euclidean $D3$-brane instanton ($ED3$), wraps the rigid blow-up cycle $\tau_3$. We assume that the tadpole-cancelation conditions can be satisfied with this choice by an appropriate set of background fluxes.

    In this case open-string loops arising from the branes $D7_1$ and $D7_2$ generate 1-loop corrections to the K\"{a}hler potential of the form \cite{bhp}:
    \begin{equation} \label{dKgs}
    \delta K_{(g_{s})} = \frac{g_s \left(\mc{C}_1^{\KK} t_1
    +\mc{C}_2^{\KK} t_2\right)}{\vo}+
    \frac{\mc{C}_{12}^\ssW}  {\vo t_2},
    \end{equation}
    where $\mc{C}_i^{\KK}$, $i=1,2$, and $\mc{C}_{12}^\ssW$ are constants which depend on the complex structure moduli. We restrict ourselves to natural values for these constants, $\alpha\sim \mc{C}_i^{\KK}\sim \mc{C}_{12}^\ssW \sim \mc{O}(1)$ and $g_s\lesssim 0.1$. In what follows we find that this scenario leads to the `small hierarchy' (SH) case described above.

 \item{\em $D7$s wrapping $\tau_3$ and $ED3$ wrapping $\tau_1$:} In this case, because there are no $D7$-branes wrapping either $\tau_1$ or $\tau_2$, the open string loop correction $\delta K_{(g_s)}$ is independent of $\tau_1$. However loops of closed Kaluza-Klein strings might still introduce a dependence on the K3 or $T^4$ divisor $\tau_1$, but, as we shall argue later on, they are expected to be suppressed, and so we shall neglect them.
\end{enumerate}

\subsubsection*{Non-perturbative corrections}

Smaller than all loop corrections are non-perturbative effects, that are typically exponentially small in the small dimensionless expansion parameters. These get swamped by perturbative corrections in $K$, but dominate the corrections to $W$ since loop corrections to this are forbidden by non-renormalisation theorems \cite{NRtheorems}. The typical corrections to $W$ that arise in this way have the form
\be \label{dWfromf}
 \delta W \simeq A \, e^{-2 \pi a f} \,,
\ee
where $f$ is the appropriate holomorphic gauge coupling function for the relevant strongly interacting sector and $a$ is a constant.

A similar story applies to the gauge coupling functions themselves, although these can receive perturbative corrections, if only at one loop. At the non-perturbative level the gauge coupling function, $f_1$, for a particular gauge group can receive non-perturbative contributions from those of another gauge groups
\be \label{df1fromf2}
 \delta f_1 \simeq A_f + B_f \, e^{-2 \pi b f_2} \,,
\ee
where $A_f$ and $B_f$ are calculable constants. In specific situations the leading dependence of $W$ on $f_2$ may be through the `poly-instanton' contribution of eq.~\pref{df1fromf2} to eq.~\pref{dWfromf}, rather than from the direct instanton correction obtained by using $f_2$ directly in eq.~\pref{dWfromf} \cite{POLY}.

Consider, for instance, the two scenarios discussed above for the loop corrections:

\begin{enumerate}
 \item {\em $D7$s wrapping $\tau_1$ and $\tau_2$; $ED3$ wrapping $\tau_3$:} In this case the $ED3$ generates a non-perturbative contribution to the superpotential of the form:
    \begin{equation} \label{Wnp}
    W = W_0 + A_3 \,e^{-2\pi T_3},
    \end{equation}
    where the tree-level superpotential $W_0$ and the threshold effect $A_3$ are $T_3$-independent constants once the $S$ and $U$-moduli are fixed and integrated out.

 \item{\em $D7$s wrapping $\tau_3$ and $ED3$ wrapping $\tau_1$:} Assuming the gauge sector of the $D7$s to involve two gauge group factors that independently condense, one expects to generate a racetrack superpotential:
    \be
    W = W_0+A \,e^{-a_3 T_3}-B \,e^{- b_3 T_3}.
    \label{racetrack}
    \ee
    On the other hand, as discussed in more detail below, the $ED3$ on $\tau_1$ can generate poly-instanton corrections \cite{POLY} to the superpotential (\ref{racetrack}) of the form:
    \bea \label{Wpoly}
    W &=& W_0 + A \,e^{-a_3\left(T_3+C_1 e^{-2\pi T_1}\right)}-B \,e^{-b_3\left(T_3+C_2 e^{-2\pi T_1}\right)} \nn\\
    &\simeq& W_0+ A \, e^{-a_3 T_3} \Bigl( 1 - a_3 C_1 e^{-2\pi T_1} \Bigr)
    - B \, e^{-b_3 T_3} \Bigl( 1 - b_3 C_2 e^{-2\pi T_1}\Bigr) \,.
    \eea
\end{enumerate}
It is this setup that leads to the huge hierarchy between $L$ and $l$ appropriate to the `large hierarchy' (LH) case described above.

We now discuss these two different cases in somewhat more detail.

\begin{figure}[ht]
\begin{center}
\epsfig{file=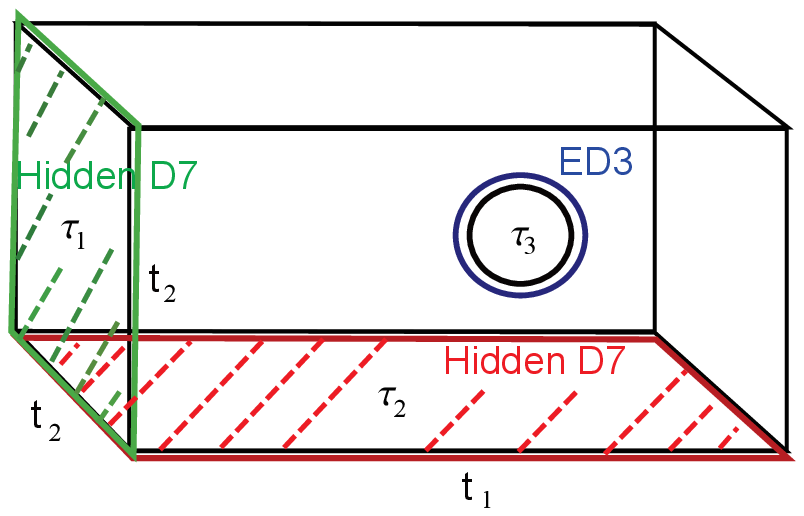, height=60mm,width=70mm}
\epsfig{file=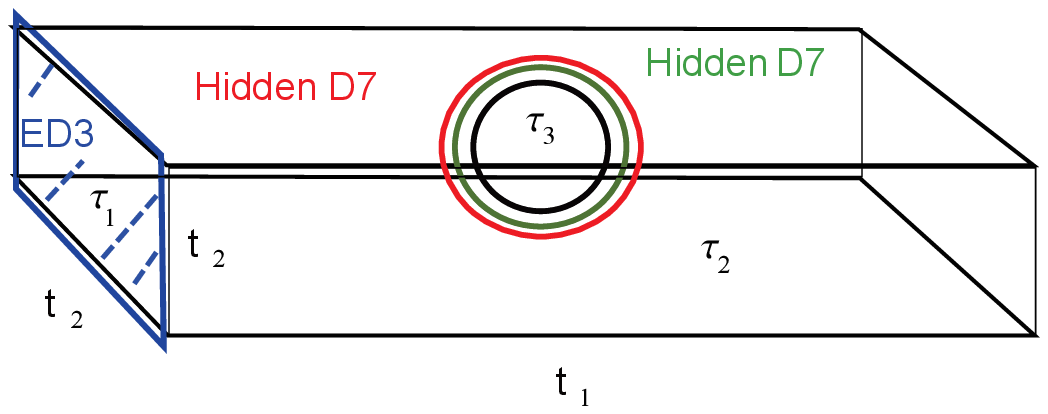, height=60mm,width=70mm}
\caption{Pictorial view of the small hierarchy case (left panel) and large hierarchy case (right panel).}
 \label{Pictures}
\end{center}
\end{figure}

\subsection{Small hierarchy} \label{bene}

Using the leading order --- tree level, eq.~(\ref{Kalpha}) --- K\"ahler potential and the leading order nontrivial superpotential --- non-perturbative, eq.~ (\ref{Wnp}) --- in the `small hierarchy' scenario gives rise to the following F-term scalar potential (after minimising with respect to the axion, $\psi_3 = \hbox{Im} \, T_3$)
\begin{equation}
 V_\ssF = \frac{32\pi^2 A_3^2}{3\alpha\gamma}  \frac{\sqrt{\tau_3}}{\vo}
 e^{-4\pi\tau_3} -8\pi W_0 A_3 \frac{\tau_3}{\vo^2}
 e^{-2\pi\tau_3} +\frac{3 \xi W_0^2}{4 g_s^{3/2}\vo^3} \,.
 \label{Vpot}
\end{equation}
Notice that $V_\ssF$ depends only on $\tau_3$ and the particular combination of $\tau_1$ and $\tau_2$ that corresponds to the overall volume, $\vo$. Consequently one combination of $\tau_1$ and $\tau_2$ parameterises a flat direction (within this approximation), while the potential (\ref{Vpot}) fixes the other two fields, $\tau_3$ and $\vo$,
\begin{equation}
 \langle \tau_3\rangle = \frac{1}{g_s}\left(
 \frac{\xi}{2 \alpha \gamma } \right)^{2/3},\,\,\,\,\,\,
 \langle\vo\rangle = \left( \frac{ 3 \alpha \gamma }{8\pi A_3}
 \right) W_0  \sqrt{\langle\tau_3\rangle}
  e^{2\pi\langle \tau_3\rangle },  \label{x}
\end{equation}
where we assume $\xi\propto (h_{2,1}-h_{1,1})>0$ in order to have a sensible solution for $\langle\tau_3\rangle$.

This reveals the LARGE-volume magic: the minimum is generically at exponentially large volume, since $\langle \tau_3 \rangle \sim \cO(1/g_s)$ and $\langle \cV \rangle \propto e^{ 2\pi/g_s }$, without fine-tuning the background fluxes, {\em i.e.} $W_0\sim\mathcal{O}(1)$. For example, the following illustrative numerical choices for the various underlying parameters,
\begin{eqnarray}
 &&\lambda_1 = \lambda_2 =1 \, \quad  \left( \hbox{and so} \;\; \alpha=0.5, \gamma=0.385\right); \nn\\
 &&g_s=0.1, \xi=0.47 \quad \left( \hbox{and so} \;\; \langle \tau_3 \rangle= 11.42 \right) \,, \notag \\
 \hbox{and} \quad && W_0 = A_3 = 1 \quad \hbox{so} \quad \langle\vo\rangle=1.15 \cdot 10^{30}. \notag
\end{eqnarray}
lead to values of $\cV$ large enough to allow $M_s \sim 1$ TeV.

The flat direction in the $(\tau_1,\tau_2)$-plane is lifted once corrections to the above choices for $K$ and $W$ are included, and first arise once loop corrections are included in the K\"{a}hler potential.

\subsubsection*{String loop corrections} \label{5}

The open-string loop corrections to $K$ in this model are estimated in eq.~(\ref{dKgs}), with the first term coming from the tree-level exchange of closed strings carrying Kaluza-Klein momentum between the $D7_1$ or $D7_2$ branes and spacetime filling $D3$-branes (whose presence is required in general due to tadpole cancelation conditions). The second term similarly arises due to the tree-level exchange of winding strings between the intersecting $D7$ stacks, $D7_1$ and $D7_2$.

Inserting the corrections of eq.~(\ref{dKgs}) into the scalar potential gives the sub-leading contribution to $V_\ssF$ in inverse powers of $\cV$. Because they are perturbatively small they do not ruin the minimum, (\ref{x}), but they can lift the flat direction of the lowest-order solution. The potential turns out to take the form \cite{Loops},
\begin{eqnarray}
 \delta V_{\left( g_s\right)}&=&\left[
 \left( g_s \mc{C}_1^{\KK} \right)^2 K^0_{1\bar 1}
 + \left( g_s \mc{C}_2^{\KK} \right)^2 K^0_{2\bar 2}
 - 2 \frac{\mc{C}_{12}^\ssW}{\vo t_2}\right] \frac{W_0^2}{\vo^2} \notag \\
 &=& \left(\frac{\mc{A}}{\tau_1^2}
 - \frac{\mc{B}}{\vo\sqrt{\tau_1}} +\frac{\mc{C}\tau_1}
 {\vo^2}\right)\frac{W_0^2}{\vo^2} \,,
 \label{74}
\end{eqnarray}
where
\begin{eqnarray}
 \mc{A} &=& \left(g_{s} \,\mc{C}_{1}^{KK}\right) ^2>0, \nonumber\\
 \mc{B} &=& 2 \, \mc{C}_{12}^{W}\lambda_1^{-1/2}
 = 4\alpha \mc{C}_{12}^{W}, \label{GREAT} \\
 \mc{C} &=& 2\,\left(\alpha g_{s}\, \mc{C}_{2}^{KK}\right)^2>0 \,. \nonumber
\end{eqnarray}
Notice that $\mc{A}$ and $\mc{C}$ are both positive (and suppressed by $g_s^2$) but $\mc{B}$ can take either sign.

\subsubsection*{Fibre stabilisation}

The structure of $\delta V_{(g_s)}$ makes it very convenient to use $\tau_1$ to parameterise the flat direction. Minimising $\delta V_{(g_s)}$ with respect to $\tau_1$ at fixed $\vo$ and $\tau_3$ gives
\be
  \frac{1}{\langle\tau_1\rangle^{3/2}} = \left( \frac{\mc{B}}{8 \mc{A} \vo} \right)
  \left[ 1 + (\hbox{sign} \, \mc{B}) \sqrt{1 + \frac{32 \mc{A}\mc{C}}{\mc{B}^2}}
  \right] \label{tau1soln1} \,,
\ee
which, when $32 \mc{A}\mc{C} \ll \mc{B}^2$ --- or equivalently $g_s^2 \ll \mc{C}_{12}^W/(2\mc{C}_1^{KK}\mc{C}_2^{KK})$ --- reduces to:
\be
  \langle\tau_1\rangle \simeq \left(-\frac{\mc{B} \vo}{2\mc{C}} \right)^{2/3}
  \hbox{if $\mc{B}<0$}
  \,\,\,\,\,\, \hbox{or} \,\,\,\,\,\,
  \langle\tau_1\rangle \simeq \left(\frac{4\mc{A} \vo}{\mc{B}} \right)^{2/3}
  \hbox{if $\mc{B}>0$} \,. \label{tau1soln2}
\ee
In order to have sensible solutions we must require either $\mc{C}> 0$ (if $\mc{B}<0$) or $\mc{A} > 0$ (if $\mc{B}>0$), a condition that is always satisfied (see (\ref{GREAT})).

The proportionality $\tau_1 \propto \cV^{2/3}$ shows that this modulus also naturally stabilises at hierarchically large values, $\tau_2 > \tau_1 \gg \tau_3$, without making unusual choices for the parameters in the potential. A useful illustrative benchmark choice of parameters is
\begin{eqnarray}
 \mc{C}^{\KK}_1 &=& \mc{C}_2^\KK = 0.1, \; \mc{C}^\ssW_{12} = 5 \,,
 \left(\hbox{which imply} \;\; \mc{A}=10^{-4},\, \mc{B}=10,\, \mc{C}=5\cdot 10^{-5}\right), \notag \\
 &&\hbox{in which case} \quad \langle\tau_1\rangle=1.3 \cdot 10^{17} \quad
 \hbox{and} \quad
 \langle\tau_2\rangle=3.2 \cdot 10^{21}. \notag
\end{eqnarray}
This construction is essentially identical to the one used in \cite{fiberinfl} to derive an inflationary model, whose inflaton is $\tau_1$, although inflationary applications require smaller values for the volume, $\vo \sim 10^3$, in order to provide observable density fluctuations. Because of this smaller volume the modulus $\tau_1$ is minimized at smaller values, and $M_s$ is of order the GUT-scale.

Unfortunately this framework does {\em not} allow a sufficiently large hierarchy between $\tau_1$ and $\tau_2$ without also building in a large hierarchy into the parameters of the potential. In particular, to get smaller values for $\langle\tau_1\rangle$ --- and so also a larger hierarchy between $L \simeq \sqrt{t_1} M_s^{-1} \simeq \left(\tau_2^{1/2}/\tau_1^{1/4}\right) M_s^{-1}$ ($\sim 3 \times 10^6 \; M_s^{-1}$ in the above example) and $l\simeq \tau_1^{1/4} M_s^{-1}$ ($\sim 2 \times 10^4 \; M_s^{-1}$ in the example) --- at fixed $\langle\vo\rangle \sim 10^{30}$, requires pre-tuning a very small hierarchy into the values of $g_s$ and the coefficients of the loop-corrected potential.

\subsection{Large hierarchy}

The brane set-up chosen above in the `large hierarchy' example is meant to ensure the dominance of poly-instanton corrections to the superpotential, of the form (\ref{Wpoly}). Before explaining their crucial r\^ole in fixing the fibre divisor at small values, let us present a brief description of this new kind of non-perturbative effect.

\subsubsection*{Poly-instanton corrections}

The authors of \cite{POLY} noticed that the action of a string instanton can receive non-perturbative corrections from another instanton wrapping a different internal cycle. They considered a Type I $T^6/\left(\mbb{Z}_2\times\mbb{Z}_2\right)$ compactification with an Euclidean $D1$ instanton ($ED1$) wrapping the $\mbb{P}^1$ base which gives rise to a single instanton contribution to the superpotential, and another $ED1$ wrapping a $T^2$ which
induces an instanton correction to the instanton action on $\mbb{P}^1$. The $ED1$ on $T^2$ does not contribute as a single instanton in $W$
due to the presence of two extra fermionic zero modes which are Wilson line modulini with the corresponding bosonic partners that are projected out.

The corresponding Type IIB version involves two internal 4-cycles, $\Sigma_i$ and $\Sigma_j$,
wrapped respectively by the Euclidean $D3$-brane instantons $ED3_i$ and $ED3_j$.
The function $f_i$ can get instanton corrections from $ED3_j$ of the form:
\begin{equation}
 f_i= \hbox{Vol}(\Sigma_i) + h(F_i)S + f_i^{1-loop}(U)
 +A_j(U)\,e^{-2\pi {\rm Vol}(\Sigma_j)},
 \label{fA}
\end{equation}
where $h(F_i)$ is a function of the world-volume flux $F_i$ on $\Sigma_i$, and the 1-loop correction $f_i^{1-loop}$ can only depend on the complex structure moduli $U$. Now the fact that the instantonic action of $ED3_i$, which we shall call $S_i$,
is related to the gauge kinetic function $f_i$ on fictitious $D7$-branes wrapping $\Sigma_i$
implies that the instanton action $S_i$ gets non-perturbative corrections which look like:
\begin{equation}
 S_i\to S_i+e^{-S_j}.
\end{equation}
In turn, the $\mc{N}=1$ superpotential takes the form:
\begin{equation}
 W=W_0+A_i\, e^{-2\pi\left(T_i+C_j e^{-2\pi T_j}\right)}.
 \label{WPOLY}
\end{equation}
The topological conditions on $\Sigma_j$ such that $ED3_j$
does not contribute to $W$ as a single instanton but just as a poly-instanton correction to $ED3_i$
have not been worked out in detail yet for the T-dual Type IIB version of the Type I computation of \cite{POLY}.
Given that determining the
details of the Type IIB origin of these poly-instanton corrections is beyond the scope of our paper,
we shall take a very phenomenological approach and assume that these effects get generated
if we identify $\Sigma_i$ with the $\tau_3$-cycle and $\Sigma_j$ with the $\tau_1$-cycle.
We shall then also assume that the gauge group on $\tau_3$ can be broken into two separate gauge groups which separately undergo gaugino condensation, so that the superpotential (\ref{WPOLY}) takes the form (\ref{Wpoly}). Then the $ED3$ on $\tau_1$ gives rise to non-perturbative corrections to the gauge kinetic functions of both condensing gauge theories.

We finally point out that it is not so unlikely that an $ED3$ on the fibre divisor gives rise
to the kind of poly-instantons corrections we need. In fact,
in the case of Type IIB orientifolds with O7/O3-planes,
the two Wilson line modulini of the Type I picture get mapped to
elements of $H^{1,0}_+(ED3)$ or $H^{2,0}_+(ED3)$ which correspond,
respectively, to Wilson line and deformation modulini.
Given that the fibre of our compactification manifold can be either a
K3 (for which $h^{2,0}=1$ and $h^{1,0}=0$) or a $T^4$ divisor
(for which $h^{2,0}=1$ and $h^{1,0}=2$), we have both Wilson line
and deformation modulini.

Let us analyse the two different situations a bit more in detail:
\begin{enumerate}
\item \emph{$ED3$ wrapping a K3 fibre}: In this case we need
$h^{2,0}_+(K3) = 1$ and $h^{2,0}_-(K3) = 0$, implying that the $ED3$ is on
top of the $O7^-$-plane giving an $Sp(2)$ instead of an $O(1)$ instanton
which does not contribute to the superpotential.
However the $ED3$ might contribute to $W$ if we magnetise the instanton since
the fluxes (both background and word-volume fluxes) may play a crucial r\^ole to
lift the universal zero-modes of an $Sp(2)$ instanton
giving rise to a contribution to the superpotential \cite{Lifting}.
A more exotic option would be to consider an $O7^+$ instead of an $O7^-$-plane
in which case an $ED3$ sitting on top of the $O7^+$-plane would correctly yield
an $O(1)$ instanton. However in this case it is likely that the bosonic
partners get projected out only if $h^{2,0}_+(K3) = 0$ and
$h^{2,0}_-(K3)=1$ implying that we have instead a $U(1)$ instanton
(which could also still contribute to $W$ in the case of a fluxed instanton).
Finally, it would also be interesting to consider the case with an $O3$-plane.

\item \emph{$ED3$ wrapping a $T^4$ fibre}: This case seems more promising
due to the presence of Wilson line modulini since $h^{1,0}(T^4) = 2$.
Here the deformations can be fixed by the background fluxes that we have turned on
to fix the dilaton and the complex structure moduli. Then,
taking the appropriate orientifold projection,
we would be left over just with Wilson line modulini that
should give rise to poly-instanton corrections.
\end{enumerate}
Hence we conclude that it is not unreasonable to assume that the
poly-instantons get generated either for a fluxed instanton on a K3 fibre
or for an instanton on a $T^4$ fibre with deformations fixed by the background fluxes. It would be interesting, but beyond the scope of this article, to have a
concrete realisation of poly-instantons in these IIB compactifications.

\subsubsection*{Modulus stabilisation}

To compute the stabilised values of the moduli we set $\alpha=\gamma=1$ (for simplicity) and trade $\tau_2$ for the volume $\vo$ using $\tau_2 = \left(\vo + \tau_3^{3/2}\right)/\sqrt{\tau_1}$. The $\mc{N}=1$ F-term scalar potential at leading order in a large volume expansion then reads (writing $T_i=\tau_i + i \, \psi_i$, $\forall i$),
\be
 V_\ssF = V_{\mathcal{O}(\vo^{-3})} +
 V_{\mathcal{O}(\vo^{-3-p})},\text{ \ \ with \ \ }p>0,
\ee
where $\cO(\vo^{-n})$ counts both explicit powers of $1/\vo$ and powers of $e^{-a_3 \tau_3} \propto 1/\vo$. Explicitly,
\begin{eqnarray}
 V_{\mathcal{O}(\vo^{-3})}&=&\frac{8 \sqrt{\tau_3} \left[A^2 a_3^2 e^{-2 a_3 \tau_3}-2 A B a_3 b_3 e^{- a_3 \tau_3-b_3 \tau_3}
 \cos (a_3 \psi_3- b_3 \psi_3) +B^2 b_3^2 e^{-2 b_3 \tau_3}\right]}{3 \vo} \notag \\
 && \qquad +\frac{4 W_0 \tau_3  \left[A a_3 e^{-a_3\tau_3} \cos(a_3\psi_3)-B b_3 e^{-b_3\tau_3}
 \cos (b_3\psi_3)\right]}{\vo^2}+\frac{W_0^2 \hat{\xi} }{\vo^3}, \notag
\end{eqnarray}
and at leading order (where, to be as general as possible, we write the exponential term $e^{-2\pi T_1}$ as $e^{-c T_1}$)
\begin{gather*}
 V_{\mathcal{O}(\vo^{-3-p})} = - \frac{16\sqrt{\tau_3} e^{-c\tau_1}}{3\vo}
 \Bigl( A^2 a_3^3 C_1 e^{-2 a_3\tau_3}+B^2 b_3^3 C_2 e^{-2 b_3\tau_3} \Bigr)
 \cos (c\psi_1) \\
 +\frac{4 W_0 e^{-c\tau_1}}{\vo^2} \Bigl[ B b_3
 C_2\left( b_3\tau_3+c\tau_1\right) e^{-b_3\tau_3}
 \cos (b_3\psi_3+c\psi_1) -A a_3 C_1\left(
 a_3\tau_3+c\tau_1\right) e^{-a_3\tau_3}\cos (a_3\psi_3+c\psi_1)\Bigr]  \\
 + \frac{16 A B a_3 b_3\sqrt{\tau_3} e^{-a_3\tau_3-b_3\tau_3-c\tau_1}}{3\vo}
 \Bigl[ b_3 C_2\cos (a_3\psi_3-b_3\psi_3-c\psi_1)+a_3 C_1\cos (a_3\psi_3-b_3\psi_3+c\psi_1)\Bigr] +P,
\end{gather*}
where $P$ is a $\tau_1$-independent piece (which we neglect from now on since our interest is in the minimisation of $\tau_1$).

We start by minimising $V_\ssF$ with respect to the axion $\psi_3 = \hbox{Im} T_3$, whose leading appearance in the potential is dominated in the term of order $\vo^{-3}$. The relevant derivatives are:
\begin{eqnarray}
 \frac{\partial V_\ssF}{\partial \psi_3}&=&\frac{4 W_0 \tau_3  \left[B b_3^2 e^{-b_3 \tau_3} \sin(b_3 \psi_3)-A a_3^2 e^{-a_3\tau_3}
 \sin (a_3 \psi_3)\right]}{\vo^2} \notag \\
 && \qquad +\frac{16 A B a_3 b_3 \sqrt{\tau_3} (a_3-b_3) e^{- a_3\tau_3-b_3\tau_3}
 \sin (a_3\psi_3-b_3\psi_3)}{3 \vo}, \notag \\
 \frac{\partial^2 V_\ssF}{\partial \psi_3^2}&=&\frac{4 W_0 \tau_3 \left[B b_3^3 e^{-b_3\tau_3} \cos (b_3\psi_3)-A a_3^3 e^{-a_3\tau_3}
 \cos (a_3\psi_3)\right]}{\vo^2} \notag \\
 &&\qquad +\frac{16 A B a_3 b_3 \sqrt{\tau_3} (a_3-b_3)^2 e^{-a_3\tau_3-b_3\tau_3}
 \cos (a_3\psi_3-b_3\psi_3)}{3 \vo}. \notag
\end{eqnarray}
Notice that ${\partial V_\ssF}/{\partial \psi_3}$ automatically vanishes at $\psi_3 = 0$, and this is a minimum if
\be
 \left. \frac{\partial^2 V_\ssF}{\partial \psi_3^2} \right|_{\psi_3=0}
 =\frac{4 W_0 \tau_3 \left(B b_3^3 e^{-b_3\tau_3}
 -A a_3^3 e^{-a_3\tau_3}\right)}{\vo^2}
 +\frac{16 A B a_3 b_3 \sqrt{\tau_3} (a_3-b_3)^2
 e^{-a_3\tau_3-b_3\tau_3}}{3 \vo} \,,
 \label{V2}
\ee
is positive.

Assuming this to be true we can (classically) integrate out $\psi_3$ by setting it to zero, leaving the residual potentials
\begin{eqnarray}
 V_{\mathcal{O}(\vo^{-3})}&=&\frac{8 \sqrt{\tau_3} \left(A^2 a_3^2 e^{-2 a_3 \tau_3}-2 A B a_3 b_3 e^{- a_3 \tau_3-b_3 \tau_3}
 +B^2 b_3^2 e^{-2 b_3 \tau_3}\right)}{3 \vo} \notag \\
 && \qquad +\frac{4 W_0 \tau_3  \left(A a_3 e^{-a_3\tau_3}-B b_3 e^{-b_3\tau_3}\right)}{\vo^2}+\frac{W_0^2 \hat{\xi} }{\vo^3},
\end{eqnarray}
and
\begin{eqnarray}
 V_{\mathcal{O}(\vo^{-3-p})} &=& \left\{ -\frac{16\sqrt{\tau_3}
 e^{-c\tau_1}}{3\vo}\Bigl[A^2 a_3^3 C_1 e^{-2 a_3 \tau_3}
 +B^2 b_3^3 C_2 e^{-2 b_3 \tau_3}-A B a_3 b_3
 e^{-a_3\tau_3-b_3\tau_3} \left(a_3 C_1 +b_3 C_2
 \right) \Bigr]\right. \notag \\
 &&\qquad \left. +\frac{4 W_0 e^{-c\tau_1}}{\vo^2}
 \Bigl[B b_3 C_2 \left(b_3\tau_3+c\tau_1\right) e^{- b_3 \tau_3}
 - A a_3 C_1 \left(a_3\tau_3+c\tau_1\right)e^{-a_3 \tau_3}
 \Bigr] \right\}\cos(c\psi_1) \,. \nn\\
\end{eqnarray}
The case of a single gaugino condensate on $\tau_3$ with polyinstanton corrections can be easily recovered setting $A$ or $B$ to zero.

Let us now evaluate $\tau_3$ at its minimum. Notice that this is determined by the dominant $\cO(\vo^{-3})$ term, but because this is independent of $\tau_1$ the resulting potential for this field is found by evaluating $V_{\cO(\vo^{-3-p})}$ at the resulting minimum. We first do so dropping all sub-dominant powers of $1/(a_3 \tau_3)$ and $1/(b_3 \tau_3)$, and find a potential for $\tau_1$ whose minimum lies at small moduli, and so lies outside the domain of validity of our approximations. We then show (for racetrack superpotentials) how a legitimate minimum can be found once we include subdominant contributions.

\medskip\noindent {\bf {\em A false start:}}
Dropping sub-dominant powers of $1/(a_3\tau_3)$ and $1/(b_3\tau_3)$, the vanishing of $(\partial/\partial \tau_3) V_{\mathcal{O}(\vo^{-3})}$ implies
\be
 A^2 a_3^3 e^{-2 a_3\tau_3}+B^2 b_3^3 e^{-2 b_3 \tau_3}-A B a_3 b_3 (a_3+b_3) e^{-a_3 \tau_3-b_3\tau_3}
 =\frac{3 W_0 \sqrt{\tau_3} \left(B b_3^2 e^{-b_3 \tau_3}-A a_3^2 e^{-a_3 \tau_3}\right)}{4 \vo}
\label{mintau3}
\ee
Writing $a_3=b_3+m$, eq.~(\ref{mintau3}) reduces to
\be
 e^{- b_3 \tau_3}=\frac{3 W_0 \sqrt{\tau_3}}{4 Z \vo},
 \label{mint3}
\ee
with
\be
 Z := B b_3 -A (b_3+m) e^{-m \tau_3} \,.
\label{Z}
\ee
In addition the condition (\ref{V2}) takes the form
\be
 B b_3- A (b_3+ m) e^{-m \tau_3}>0,
\label{cond}
\ee
which implies $Z>0$. In the special case $m=0$ ({\em i.e.} when $a_3=b_3$) $Z >0$ reduces to $B > A$ (since $b_3 > 0$). In the case of a single exponential ($A=0$) we have $Z=B b_3$.

Writing $C_2=C_1+n$, the scalar potential for $\tau_1$ then becomes
\begin{gather}
 V_{\mathcal{O}(\vo^{-3-p})} =\left\{C_1\left[-\frac{16 \sqrt{\tau_3}e^{-2 b_3 \tau_3}}{3\vo}\left[
 B^2 b_3^3 -A B b_3 (b_3+m)\left(2 b_3 +m\right) e^{-m\tau_3}+A^2 (b_3+m)^3 e^{-2 m \tau_3}\right]\right.\right. \notag \\
 \left. +\frac{4 W_0 e^{- b_3 \tau_3}}{\vo^2}\left[B b_3 \left(b_3\tau_3+c\tau_1\right)
- A (b_3+m) \left((b_3+m)\tau_3+c\tau_1\right)e^{-m\tau_3} \right] \right] \label{VO4f} \\
 \left.+ n \left[-\frac{16 B b_3^2 \sqrt{\tau_3} e^{-2 b_3 \tau_3} \left(B b_3
 - A  (b_3+m)  e^{-m \tau_3}\right)}{3 \vo}
 +\frac{4 W_0 B b_3 e^{-b_3 \tau_3}}{\vo^2}\left(b_3\tau_3+c\tau_1\right)\right]\right\}e^{-c\tau_1} \cos (c\psi_1) \notag
\end{gather}
which takes the form
\be
 V_{\mathcal{O}(\vo^{-3-p})} =\frac{3 W_0^2 \sqrt{\tau_3}}{Z\vo^3}\left(r_1 c \tau_1 + r_2 b_3\tau_3 \right)e^{-c\tau_1} \cos (c\psi_1) \notag
\label{VO4finale}
\ee
once eq.~(\ref{mint3}) is used. The quantities $r_1$ and $r_2$ evaluate to
\be
 r_1 = C_1 Z+n B b_3 \quad \hbox{and} \quad r_2 = 0 \,.
\ee
The final leading-order potential for $\tau_1$ is therefore
\be
 V_{\mathcal{O}(\vo^{-3-p})}=\frac{\beta }{\vo^3} \,c\tau _1
 e^{-c \tau _1} \cos\left(c \psi _1\right),
\label{V1}
\ee
with $\beta$ an $\mc{O}(1)$ constant which does not depend on $\tau_1$.
Unfortunately, the global minimum of this potential is at $c \langle \psi_1 \rangle = \pi$
and $c \langle \tau_1 \rangle = 1$, which lies outside the large-modulus regime
where we trust our effective field theory treatment.
In particular, in the case of interest $c = 2\pi$ since the fibre divisor
is wrapped by an Euclidean $D3$-brane instanton,
leading to too small a value for $\langle \tau_1 \rangle = 1/(2\pi) < 1$.

\medskip\noindent {\bf {\em A better approach:}}
We next show that the potential can have solutions within a trustable regime provided we include the sub-leading corrections to the expression (\ref{mint3}) in powers of $1/(a_3\tau_3)$ and $1/(b_3\tau_3)$, that had earlier been dropped. It turns out that even these would not save the day if we had assumed a single-exponential superpotential, and it is for this reason that we instead started with a racetrack superpotential, as appropriate to the condensation of two gauge group factors. The racetrack form helps by allowing the sub-leading corrections to compete with the potential, eq.~(\ref{V1}). Let us see why.

The sub-leading corrections to the expression (\ref{mint3}) in a $1/(a_3\tau_3)$ and $1/(b_3\tau_3)$ expansion are given by:
\be
 e^{- b_3\tau_3}=\frac{3 W_0 \sqrt{\tau_3}}{4 Z \vo}f_{\rm corr}
\ee
where
\be
 f_{\rm corr}\equiv 1-\frac{3\epsilon}{1+m \left( \frac{1}{b_3}-\frac{B_3}{Z}\right)}\,,
 \label{MinTau3}
\ee
with
\be
 \epsilon\equiv \frac{1}{4 b_3 \tau_3}\ll 1\text{ \ \ for \ \ }b_3\tau_3\gg 1.
\ee
We notice that in the single exponential case ($m=0$), (\ref{MinTau3}) reduces to:
\be
 f_{\rm corr}=1-3\epsilon,
\ee
implying that for $b_3\tau_3\gg 1$, the corrections are always subleading.

Substituting now the new corrected result (\ref{MinTau3}) in (\ref{VO4f}), we find the corrected potential
\be
 V_{\mathcal{O}(\vo^{-3-p})} =\frac{3 W_0^2 \sqrt{\tau_3}}{Z\vo^3}f_{\rm corr}
 \left(r_1 c \tau_1+r_3 b_3\tau_3 \right)e^{-c\tau_1} \cos (c\psi_1),
\label{VO4Finale}
\ee
where this time
\be
 r_3\equiv \left[\frac{r_1}{b_3}(b_3+m)-m B(C_1+n)\right](1-f_{\rm corr}) \,,
\ee
does not vanish (though it would if $\epsilon=0$, since this implies $f_{\rm corr}=1$).

The potential (\ref{VO4Finale}) is of the form:
\be
 V_{\mathcal{O}(\vo^{-3-p})}=\frac{\beta}{\vo^3}
 \left(c \tau _1-p \,b_3\tau_3\right) e^{-c \tau _1} \cos\left(c \psi _1\right),
\label{V}
\ee
where $\beta$ is an unimportant $\mc{O}(1)$ constant while $p$ is given by:
\be
 p\equiv -\frac{r_3}{r_1}=\left[\frac{m B(C_1+n)}{r_1}-\frac{(b_3+m)}{b_3}\right](1-f_{\rm corr}) \,.
\ee
The potential (\ref{V}) admits a global minimum at
$c \langle\psi_1\rangle=\pi$ and
$c \langle \tau_1 \rangle = p \,b_3 \langle\tau_3\rangle+1\simeq p \,b_3 \langle\tau_3\rangle$,
regardless of the value of $\beta$ (which determines only the depth of the vacuum).

In the case of only a single exponential ($m=0$), $p$ becomes negative with an absolute value smaller than unity:
\be
|p|=3 \epsilon =\frac{3}{4 b_3 \tau_3 }\ll 1\,\,\, \Rightarrow \,\,\,
c \langle\tau_1\rangle = -\frac{3}{4}<0,
\ee
and so we end up in a regime where the minimum for $\tau_1$ is out of the K\"ahler cone.
However in the racetrack case it is possible to render $p$ positive and large enough
to trust the effective field theory. Consider the following illustrative, benchmark, values (with $a_3=2\pi/N_a$ and $b_3=2\pi/N_b$):
\be
W_0=B=N_b=10,\,A=0.02,\,N_a=11,\,
C_1=1,\,n=-0.4506,\,c=2\pi,\,\xi=0.7,\,g_s=0.01. \notag
\ee
These numbers yield
\be
\langle\tau_3\rangle  \simeq\frac{\left(\sqrt{2} \,\xi\right)^{2/3}}{g_s}\simeq\frac{1}{g_s}=100,
\text{ \ \ }p=0.97\simeq 1\,\,\,\Rightarrow\,\,\,\langle\tau_1\rangle\simeq \frac{\langle\tau_3\rangle}{10}=10,
\ee
with $Z=2.83>0$ and $f_{\rm corr}= 0.99$. Notice that the value $l=\langle \tau_1 \rangle \,l_s\simeq 10 \,l_s$ gives a good large-modulus approximation since corrections are controlled by\footnote{If a $D7$ brane had been wrapped on this cycle there is also a 4D understanding of why the $4 \pi$'s break our way like this. Since any gauge coupling for such a $D7$ satisfies $\tau_1 = 4\pi/g^2$, gauge loops are controlled by $(g/4\pi)^2 = 1/(4\pi \tau_1)$ and are small even if $g^2 = 4\pi/\tau_1$ is order unity. } $\alpha'/l^2 = 1/\left(4\pi^2\langle \tau_1 \rangle^{1/2}\right) \simeq 1/12\pi^2$.
Moreover, we stress the need for moderate fine tuning in the choice of the parameter $n$ in order to obtain $p$ large enough to trust the effective field theory description.

The overall volume in this case evaluates to the extreme case of TeV-scale strings
\be
 \vo\simeq 5.2\times 10^{28}\,\,\,\Rightarrow\,\,\,M_s=\frac{M_p}{\sqrt{4\pi\vo}}\simeq 3\,\text{TeV}.
\ee

Some comment is required as to why we choose $g_s$ as small as $1/100$. This is driven by the interplay of the two conditions:
\begin{enumerate}
 \item $a_3$ should be close enough to $b_3$ to allow $p$  to be sufficiently large;

 \item $\vo$ should be large enough, $\vo\sim 10^{30}$, to obtain TeV-scale strings.
\end{enumerate}
Given that the volume goes like $\vo \sim e^{b_3 / g_s}$, if $g_s=0.1$ then $b_3$ has to be $b_3=2 \pi$. However with such a large $b_3$, $a_3$ can not be very close to $b_3$ (at most we can choose $a_3=\pi$). On the contrary for $g_s=0.01$ then $b_3$ can be $b_3=2\pi/10$, and so $a_3=2\pi/11$ can now be very close to $b_3$. In order to allow larger values of $g_s$, one should drop the phenomenological requirement of getting TeV-scale strings, or keep it but then allowing more fine-tuning in the choice of the other parameters.

To summarise: the above construction shows how poly-instanton corrections open up the possibility of achieving both a very large volume (to allow $M_s \sim 1$ TeV), and a very anisotropic shape of the compactification manifold (to allow a huge hierarchy among the sizes of the different dimensions)
\be
 d \simeq \langle\tau_3\rangle^{1/4} l_s\gtrsim l \simeq \langle \tau_1 \rangle^{1/4} l_s\sim 10^{-17} \,
 \hbox{ mm}  \ll L \simeq \langle t_1 \rangle^{1/2} l_s= \sqrt{\langle\vo\rangle/\langle\tau_1\rangle}\, l_s\sim 0.01 \, \hbox{ mm} \,.\nn
\ee

\medskip\noindent {\bf {\em Closed string loops:}}
We conclude this section by pointing out that we do not expect any $\tau_1$-dependent open-string loop correction to $K$ due
to the absence of $D7$-branes wrapping the K3 or $T^4$ divisor, so that no open strings are localised on $\tau_1$.
However there is no way to avoid by construction loops of closed Kaluza-Klein strings
which might introduce a dependence on the fibre divisor and be dangerous for our scenario
if they dominate over the tiny poly-instanton effects.

We shall now argue that this might not be the case
since the contribution of the closed-string loops to the vacuum energy
can be estimated to scale as
\be
\delta V_{(g_s)} \sim \Lambda^2\,\, {\rm STr} (M^2)\sim (M_{\KK}^{6D})^2 m_{3/2}^2\sim \frac{\tau_1}{\vo^4},
\label{Closedloops}
\ee
where we used the 1-loop Coleman-Weinberg potential \cite{ColeWbg} with a cut-off
given by the 6D Kaluza-Klein scale $M_{\KK}^{6D}=M_s/t_1^{1/2}\simeq M_P\sqrt{\tau_1}/\vo$ and $m_{3/2}\sim M_P/\vo$.
Given that the poly-instanton effects also scale as $1/\vo^4$, the contribution (\ref{Closedloops})
does not destabilise our scenario.

\section{Mass scales and low-energy spectrum}

This section identifies the mass scales of relevance to phenomenological applications, for both the large- and small-hierarchy examples. Because we explicitly stabilise the moduli we can be explicit about the spectrum of light states that are potentially relevant to low-energy physics, and how their properties are correlated with those of the higher-energy particles relevant to physics at the LHC.

The spectrum of bulk fields in these models shares the generic features of the LARGE volume scenario, with a rich variety of states predicted with masses and couplings that scale as different powers of the large volume, $\vo$. To these must be added more model-dependent predictions, including in particular a specification of precisely where observable Standard Model particles are situated.

We first very briefly remind the reader about the generic features, before turning to the more model-dependent assumptions about how the Standard Model fits in.

\subsection{Bulk mass scales}

Since factors of $2\pi$ can make a difference, we first summarise the basic scales occurring in our small- and large-hierarchy scenarios. Recall the Einstein term in the 10D Type IIB supergravity action in string frame is
\be
 S_{10D}^{(s)} \supset \frac{1}{(2 \pi)^7 \alpha'^4} \int d^{10}x \sqrt{-g^{(s)}_{10}}
 e^{-2 \phi}\mathcal{R}_{10}^{(s)},
\ee
and so the action in Einstein frame is obtained via the Weyl rescaling $g_{MN}^{(s)} = e^{\phi/2} g_{MN}^{(E)}$. In terms of $l_s = 2\pi\sqrt{\alpha'} = 1/M_s$ the 10D Planck scale therefore satisfies $M_{10D}^8 = 4\pi/l_s^8$, and so
\be
 M_{10D} = \left(4 \pi\right)^{1/8} M_s \simeq 1.4\, M_s.
\ee
Dimensionally reducing from 10D to 6D then yields\footnote{For simplicity we assume here that the geometry is not strongly warped.} the 6D Planck scale $M_{6D}^4 = (4 \pi/l_s^8) \; V_{\rm fib}$, where the volume of the fibre is $V_{\rm fib}= \int d^4 y \sqrt{g_4^{(E)}} := l^4 = \tau_1 l_s^4$, and so
\be
 M_{6D} = \left(4 \pi\tau_1\right)^{1/4} M_s = M_{10D}^2\, l.
\ee
Notice that because $M_{10D} \, l \simeq M_s \, l = \tau_1^{1/4} > 1$, we have $M_{6D} > M_{10D}$. The further dimensional reduction from 6D to 4D then yields $M_p^2 = (4\pi/l_s^8) \, V_6$, where $V_6 = \int d^6 x \sqrt{g_6^{(E)}} := \vo \,l_s^6$, and so we find
\be
 M_p = \sqrt{4 \pi \vo} \, M_s .
\ee

\subsubsection*{KK scales}

The extra-dimensional geometries of interest come with a variety of KK scales. The basic transition from 4D to a higher-dimensional description occurs at the smallest KK scale, which we've seen is set by the volume of the largest cycle,
\be
 M_{\KK}^{6D}=\frac{M_s}{t_1^{1/2}}= \frac{1}{L} \,.
\ee
Above this scale the effective description is 6-dimensional for a range of energies up to
\be
 M_{\KK}^{10D} = \frac{M_s}{\tau_1^{1/4}} = \frac{1}{l} \,,
\ee
above which the full 10 dimensions become visible. When $L \simeq l$ the transition is directly from 4D to 10D and the pattern of KK masses is broadly similar to what is expected if all six internal dimensions were roughly of the same size.

However there is an important difference between these examples and the simplest ADD-style models of large extra dimensions. This is due to the existence of small stabilised 4-cycles in the geometry, with sizes like $\tau_3 = d^4/l_s^4 \simeq 1/g_s \sim 10$, and so for which
\be
 M_{\KK}^c := \frac1d \simeq \frac{M_s}{\tau_3^{1/4}} \,.
\ee
Although counter-intuitive for those brought up using tori and spheres, the existence of such a variety of geometrical scales is generic for the more complicated geometries that naturally arise in flux compactifications.\footnote{The observation that higher-dimensional compactifications can be very rigid, and so have KK scales much larger than their volumes would indicate, has been occasionally  used by model-builders \cite{CHM}.}

Using the illustrative values given above for $\vo$, $\tau_1$ and $\tau_3$ in the large- and small-hierarchy cases, we find the numerical values listed in Table 1

\begin{table}[ht]
\label{tab:scales}
\begin{center}
\begin{tabular}{c|cccccc}
  & $M_s$ & $M_{6D}$ & $M_{10D}$ & $M_{\KK}^c$ & $M_{\KK}^{10D}$ & $M_{\KK}^{6D}$ \\
  \hline\hline
  \\ & & & & & & \vspace{-0.9cm}\\
  small hierarchy
  & $1$ TeV & $2000$ TeV & $2$ TeV & $0.5$ TeV & $50$ MeV & $0.3$ MeV \\
  \\ & & & & & & \vspace{-0.9cm}\\
  large hierarchy
  & $3$ TeV & $10$ TeV & $4$ TeV & $1$ TeV & $1$ TeV & $0.01$ eV
\end{tabular}
\end{center}
\caption{Relevant mass scales for small- and large-hierarchy examples using the numerical values for modulus sizes given in \S2.}
\end{table}

\subsubsection*{Generic moduli}

Some of the would-be moduli of the lowest-order theory are fixed by $D$-terms which generate $\mc{O}(M_s)$ masses, but others are systematically light compared with generic KK masses (and so can be described within the effective 4D theory). Many of these --- such as complex structure moduli, $U$, and the dilaton $S$ --- obtain masses from background fluxes, which from the 4D perspective generate a tree level $F$-term potential. These states generically couple with 4D Planck strength, and their potential scales like $V_{\ssF} \simeq M_s^4$, so the resulting masses are generically of order $M_s^2/M_p \simeq M_p/\cV$.

This is numerically of order $ \simeq 10^{-3}$ eV for $M_s \sim 1$ TeV, for both the large- and small-hierarchy examples. For large hierarchies the mass of these moduli is similar to the lightest KK states, $M_\KK^{6D}$, but they are parametrically light relative to all KK scales for the more conventional small-hierarchy case. Remarkably, a combination of low gravity scale and volume-suppressed interactions ensures these small masses are stable against radiative corrections \cite{uber}. The flux-induced contributions to the gravitino mass are similarly small, $m_{3/2} \simeq M_p/\cV \simeq 10^{-3}$ eV (more about this below).

Masses for the K\"ahler moduli are generically just as small, and can be even lighter in some instances, because of the no-scale structure which keeps them massless to leading order in $\alpha'$ and $g_s$. A detailed determination of their size requires diagonalising their kinetic and mass terms, and depends somewhat on the precise scenario considered (as is described below in detail). Before doing so we must first become more specific about precisely where the Standard Model degrees of freedom are located.

\subsection{Locating the Standard Model}

Particle phenomenology requires the identification of where Standard Model states arise within the model. For the remainder of this paper we suppose them to be localised on a brane, which we supposed to be an appropriate stack of magnetised $D7$s and $D3$s since these are known to be promising starting points for model building \cite{singular, magnetised}.
If so, the coupling of these fields to other light states depends crucially on where they are located in the extra dimensions and on which cycles the $D7$s wrap.

There are several considerations that can be used to guide the choice of the cycle wrapped by the SM $D7$ branes:
\begin{itemize}
\item it cannot be too large, or else the resulting gauge couplings, $g^2/4\pi = 1/\tau$, become too small;
\item it cannot be too large or else the KK excitations of SM states would have been observed;
\item its intersections with other cycles in the geometry must not destroy the dynamics that stabilises the moduli describing these other cycles.
\end{itemize}

We now argue that none of the cycles discussed so far for the fibred Calabi-Yau manifolds considered above are suitable to be wrapped by the SM branes in this way. Because the gauge coupling for fields on the brane is given by $4\pi/g^{2} = \tau_{\SM}$, the SM branes must wrap one of the relatively small cycles of the geometry to prevent having exponentially small gauge couplings. Hence $\tau_{\SM}$ cannot be $\tau_2$ in either of the fibred examples examined above. In the small-hierarchy case, it also cannot be $\tau_1$ for the same reason.

The fibre modulus $\tau_1$ can also be eliminated for the large-hierarchy scenario, but for a different reason. If the SM wraps $\tau_1$ then string loop corrections to $K$ depending on $\tau_1$ would be generated, and these would dominate over the tiny poly-instanton corrections, making this case degenerate with the former one.

The only candidate 4-cycle left is the blow-up mode $\tau_3$. However even this cycle cannot be $\tau_{\SM}$. For the small-hierarchy case this is because $\tau_3$ is already wrapped by an $ED3$. Also wrapping the SM brane around this cycle would then produce chiral intersections between the $ED3$ and the SM brane, which would induce a pre-factor for the instanton correction to $W$ that is proportional to powers of the SM chiral fields. But unbroken gauge symmetries require these fields to have vanishing VEVs, thereby removing any possible
non-perturbative contribution to the superpotential \cite{blumen}. That is,
\be
 W_{np}\sim \left(\Pi_i \Phi_i \right)e^{-2\pi T_3}=0\text{ \ \ with \ \ }\langle\Phi_i\rangle=0.
\ee

In the large-hierarchy case, on the other hand, non-perturbative corrections depend on $\tau_3$, since this cycle supports the branes containing the two condensing gauge sectors. The above incompatibility between chiral intersections and non-perturbative effects implies that $\tau_3$ also cannot support the SM brane in this case.

All roads lead to Rome: another 4-cycle is needed --- call it $\tau_4$ --- on which to wrap the SM branes, $D7_\SM$.
We now describe the two natural choices for the size of this cycle: making it large or small compared to the string scale. We find below that these differ in their implications for low-energy phenomenology.

\subsubsection*{The geometric regime}

The conceptually simplest choice places the SM on intersecting $D7$-branes wrapping an internal 4-cycle whose volume, $\tau_4 = \tau_{\SM}$, is stabilised at a value that is `geometric', in the sense of being larger (but not too much larger) than the string scale~\cite{GeomRegime}.

If $\tau_4$ is a blow-up mode that intersects $\tau_3$, it can be stabilised in the geometric regime using either $D$-terms \cite{blumen, IntersectingD}, or via string loop corrections to the K\"{a}hler potential \cite{LVSgeneral}. World-volume fluxes on $D7_{\SM}$ and on another stack of $D7$-branes, $D7_{\rm int}$, wrapped around a combination of $\tau_3$ and $\tau_4$, can then be appropriately chosen to ensure there are no chiral intersections between $D7_{\SM}$ and the $ED3$ (or the stack of $D7$-branes) that yields the non-perturbative superpotential. Instead they arise only between $D7_{\SM}$ and $D7_{\rm int}$ \cite{blumen}. In this way non-perturbative corrections to $W$ depending on $\tau_3$ do not get destroyed. In the absence of SM singlets that can get a non-vanishing VEV, $D$-terms can fix $\tau_4$; and in their presence $\tau_4$ could instead be fixed by $g_s$ corrections.

The upshot is that the volume, eq.~(\ref{hhh}), changes to:
\be
 \vo = \alpha \left( \sqrt{\tau_1}\tau_2 - \gamma \tau_3^{3/2}\right)
 \,\,\,\,\rightarrow\,\,\,\, \vo = \alpha \left[ \sqrt{\tau_1}\tau_2 - \gamma \left(c_3 \tau_3+c_4 \tau_4\right)^{3/2}\right].
 \label{volgeom}
\ee

The geometric scenario is strongly constrained by the existence of KK excitations of SM states in the 4 extra dimensions along the cycle wrapped by the SM brane. The absence of any evidence for such states \cite{NoSMKK,pdg} implies these KK modes cannot be lighter than ~ 1 TeV. The good news here is that such large KK masses are possible, despite the large overall size of the various dimensions, since small stabilised cycles can exist with $M_\KK^c \sim M_s/\tau_4^{1/4} \gg M_\KK^{10D}, \; M_\KK^{6D}$. In the present instance, because $\tau_4$ also sets the size of the gauge couplings, $\tau_4= 4\pi g_{\SM}^{-2}=\alpha_{\SM}^{-1}$, we have
\be
 M_{\KK}^{\SM} = \frac{M_s}{\tau_4^{1/4}} = \alpha_{\SM}^{1/4} M_s \,.
\ee
Because $\alpha_\SM$ is known this puts a direct lower bound on $M_s$ in this scenario.

\subsubsection*{Fractional branes at singularities}

The alternative to the geometric regime is to imagine the SM is built from fractional $D$-branes located at the singularity obtained by shrinking the blow-up mode supporting the SM brane: $\tau_{\SM}\to 0$ ~\cite{quiver}.
In this scenario $\tau_{\SM}$ cannot be any of the K\"{a}hler moduli discussed in previous sections, since these are all larger than $l_s$ by assumption. Again we need a fourth cycle, $\tau_4$, to support the SM branes. If this cycle is rigid and does not intersect any of the other cycles, then $\tau_4$ can be forced to shrink at the singularity, $\tau_4\to 0$, using $D$-terms \cite{quiver}.

This picture has two attractive features. First, because the SM branes do not wrap any cycles there are no KK modes for SM states and the natural scale for all excitations is the string scale. Second, the SM gauge coupling is unrelated to a cycle volume and is instead directly controlled by the string coupling, $g_s$.

In this case the volume, eq.~(\ref{hhh}), changes to:
\be
 \vo = \alpha \left( \sqrt{\tau_1}\tau_2 - \gamma \tau_3^{3/2}\right)
 \,\,\,\,\rightarrow\,\,\,\, \vo = \alpha \left( \sqrt{\tau_1}\tau_2 - \gamma_3 \tau_3^{3/2}- \gamma_4\tau_4^{3/2}\right) \,.
 \label{volsing}
\ee
The dependence of the K\"{a}hler potential on the K\"{a}hler moduli is understood in terms of an expansion in small $\tau_4$ (see below).

\subsection{Supersymmetry breaking}

Any realistic description of the Standard Model on a brane with a TeV string scale must include adequately large supersymmetry breaking. Since phenomenology requires no superpartners to ordinary particles almost up to the (TeV) string scale \cite{LHCSUSY}, the SM sector must not even approximately be supersymmetric.

As mentioned in the introduction, since the SM must in any case be localised on a brane when $M_s \simeq 1$ TeV, the most natural supersymmetry-breaking mechanism is to have the SM brane itself not be supersymmetric. For a non-supersymmetric brane, supersymmetry is only realised nonlinearly: a supersymmetry transformation acting on a particle state returns the same particle plus a brane-localised goldstino. This goldstino is then eaten by the gravitino once the brane is coupled to gravity in the bulk. The upshot is that there are no single-particle super-partners (like the selectron, say) for any of the known particles, and the low-energy limit is {\em not} described by the MSSM, even though supersymmetry is broken at the weak scale \cite{MSLED,SLEDpheno}. See \cite{singular} for explicit local constructions of non-supersymmetric branes.

Notice that in these scenarios supersymmetry only plays an indirect r\^ole in the hierarchy problem. Instead, the hierarchy problem is solved by having a TeV gravity scale, but with the modulus stabilisation mechanism providing the usually missing (but crucial) step of explaining why the extra dimensions are so large.

Although supersymmetry is badly broken in the SM sector, the scale of supersymmetry breaking this induces in the bulk turns out to be very small and similar in size to SUSY-breaking flux effects: $m_{3/2} \simeq M_s^2/M_p \simeq M_p / 4 \pi \vo \sim 10^{-3}$ eV. This small a breaking arises naturally because the bulk generically couples to the supersymmetry breaking sector with gravitational strength.

\subsection{Modulus spectrum and couplings: leading order}
\label{Mmc}

We next estimate the mixing and masses of K\"ahler moduli, including the fourth K\"ahler modulus, $\tau_4$, whose existence is required by the presence of the SM brane. In this section we canonically normalise the leading order kinetic terms and diagonalise the resulting mass matrix. (The next section discusses corrections to these leading results.) This amounts to finding the eigenvectors and the eigenvalues of the mass-squared matrix $\left(M^2\right)^i_{k} := K^{i\bar j}V_{\bar j k}$. Our goal is to track how these quantities scale with the small parameter $\vo^{-1}$. We quote the result for $\tau_4$ in both the geometric and singular regimes.

\subsubsection*{Geometric regime}

The diagonalisation of moduli states and a determination of their mass spectrum is worked out in some detail for the small-hierarchy geometries in \cite{ACC, CM}. The derivation for the large-hierarchy case is very similar, so we simply outline here the main results for the case when $\tau_4$ is stabilised in the geometric regime.

The transformation that canonically normalises the kinetic terms for fluctuations about the potential minimum reads \cite{CM}
\begin{eqnarray}
 \delta \tau_1 &\simeq& \sum_{i=1}^2 \omega_{1i}
 \, \delta \phi_i +\sum_{j=3}^4 \frac{\omega_{1j}}{\vo^{1/2}} \,
 \delta \phi_j \approx
 \sum_{i=1}^2 \omega_{1i} \, \delta \phi_i, \label{CANnorm1} \\
 \delta \tau_2 &\simeq& \sum_{i=1}^2 {\omega_{2i}}{\vo}
 \, \delta \phi_i +\sum_{j=3}^4 {\omega_{2j}}{\vo^{1/2}} \,
 \delta \phi_j \approx
 \sum_{i=1}^2 {\omega_{2i}}{\vo} \, \delta \phi_i, \label{CANnorm2} \\
 \delta \tau_k &\simeq& \frac{\omega_{k1}}{\vo^{n}} \, \delta \phi_1
 + \omega_{k2} \, \delta \phi_2 + \sum_{j=3}^4
 {\omega_{kj}}{\vo^{1/2}} \, \delta \phi_j
 \approx \sum_{j=3}^4 {\omega_{kj}}{\vo^{1/2}}
 \, \delta \phi_j \quad \hbox{for} \;  k=3,4, \label{CANnorm3}
\end{eqnarray}
where $n= \frac13$ for small hierarchies; and $n=p$ (with $p\simeq 1$ for our choice of parameters) in the large-hierarchy case. The constants $\omega_{ki}$ are order-unity constants. The resulting spectrum of modulus masses is
\be
 m_1 \simeq \sqrt{\frac{g_s}{4\pi}}\frac{M_p}{\vo^{(3+n)/2}} , \text{ \ \ }
 m_2 \simeq \sqrt{\frac{g_s}{4\pi}}\frac{M_p}{\vo^{3/2}}, \text{ \ \ }
 m_3 \simeq \sqrt{\frac{g_s}{4\pi}}\frac{M_p}{\vo} \text{ \ \ and \ \ }
 m_4 \simeq \sqrt{\frac{g_s}{4\pi}}\frac{M_p}{\vo^{m/2}} ,
\label{MASSE}
\ee
where $m=1$ if $\tau_4$ is fixed by $D$-terms, or $m=2$ if the SM cycle is fixed using $g_s$ corrections to $K$.

This is an interestingly complicated hierarchy of masses, that is exquisitely sensitive to the size of the extra-dimensional volume. Many of its features have simple physical interpretations. As shown in \cite{CM}, which inverts the exact form of (\ref{CANnorm1}), $\delta\phi_2$ turns out to be the particular combination of $\delta\tau_1$ and $\delta\tau_2$ that corresponds to the overall volume, whereas $\delta\phi_1$ is a direction orthogonal to this that is fixed only at sub-leading order --- either by string loops in the small-hierarchy case or by poly-instanton corrections for the large hierarchy. It is because $\delta \phi_1$ first receives its mass at higher order that makes it systematically lighter than $\delta\phi_2$, as can be seen from (\ref{MASSE}).

The volume-scaling of the canonical normalisation of the small blow-up modes (\ref{CANnorm3}) can also be understood from a geometric point of view. Each canonically normalised field, $\delta\phi_k$, $k=3,4$, mostly overlaps a combination of the two intersecting blow-up modes with a power of $\vo^{1/2}$. The next mixing in a large volume expansion is with $\delta\phi_2$, which corresponds to the volume mode. The $\mc{O}(1)$ mixing between the two blow-up modes is due to their non-vanishing intersection, while the suppression with respect to the mixing with $\delta\phi_2$, reflects the locality of the two blow-up modes within the Calabi-Yau volume. Finally the mixing with the other modulus $\delta\phi_1$ is further suppressed by a power of $\vo^{-n}$ due to the fact that the shape of the lagrangian in the direction of the ($\tau_1 - \tau_2$)-plane orthogonal to $\vo$ is only fixed at sub-leading order.

For later purposes what is important is that these are {\em extremely} small masses when $M_s$ is in the TeV range. In particular, $m_3 \simeq 10^{-3}$ eV, $m_2 \simeq 10^{-18}$ eV and $m_1\simeq 10^{-32}$ eV when the benchmark numbers (for $n=p=1$) of previous sections are used. These correspond to the macroscopic wavelengths: $m_3^{-1}\simeq 100$ $\mu$m; $m_2^{-1} \simeq 10^{11}$ m --- of order the Earth-Sun distance; and $m_1^{-1} \simeq 10^{25}$ m $\sim 300$ Mpc --- of order the current Hubble scale.

Because these leading contributions to masses are so light, the danger is that they are dominated by nominally subdominant effects. We examine this in the next section, and find that some get significant contributions from loops but (remarkably) others do not.

\medskip\noindent{{\em Couplings:}}

\medskip\noindent
Before turning to loops we first examine the size of the couplings between these moduli and states (like the SM) localised on the branes. As an estimate of these couplings we work out the $\cV$-dependence of the interaction
\be
 \cL_{\Sigma_i} = \frac{\zeta_i}{M_p} \,\delta \phi \, F_{\mu\nu}^{(i)} F^{\mu\nu}_{(i)} \,,
\ee
between fluctuations in these moduli and gauge bosons living on the various cycles, $\tau_i$,  $i=1,2,3,4$ \cite{CM}. To this end recall that gauge bosons only live on $\tau_1$, $\tau_2$ and $\tau_4$ in the small-hierarchy case, but only on $\tau_3$ and $\tau_4$ in the large-hierarchy example.

The $\vo$-dependence of the resulting couplings, $\zeta_i$, are summarised in Tables 2 and 3. These reveal that $\delta \phi_2$ always couples gravitationally ({\em i.e.} $\sim 1/M_p$) to all gauge sectors, while the moduli $\delta \phi_3$ and $\delta \phi_4$ always couple to fields on the SM brane with weak-interaction ({\em i.e.} higher-dimensional, as opposed to 4D, gravitational) strength. Most remarkably, the other couplings between moduli and gauge sectors can be orders of magnitude weaker than gravitational.

\begin{table}[ht]
\begin{center}
\begin{tabular}{c|cccc}
  & $\delta\phi_1$ & $\delta\phi_2$
  & $\delta\phi_3$ & $\delta\phi_4$ \\
  \hline\hline
  \\ & & & & \vspace{-0.9cm}\\
  $\zeta_{1}, \zeta_{2}$
  & $1$
  & $1$
  & $\vo^{-1/2}$
  & $\vo^{-1/2}$ \\
  \\ & & & & \vspace{-0.9cm}\\
  $\zeta_4$
  & $\vo^{-1/3}$
  & $1$
  & $\vo^{1/2}$
  & $\vo^{1/2}$
\end{tabular}
\end{center}
\caption{Modulus couplings to brane gauge bosons in the small-hierarchy geometric regime.}
\end{table}

\begin{table}[ht]
\begin{center}
\begin{tabular}{c|cccc}
  & $\delta\phi_1$ & $\delta\phi_2$
  & $\delta\phi_3$ & $\delta\phi_4$ \\
  \hline\hline
  \\ & & & & \vspace{-0.9cm}\\
  $\zeta_{3}, \zeta_{4}$
  & $\vo^{-p}$
  & $1$
  & $\vo^{1/2}$
  & $\vo^{1/2}$
\end{tabular}
\end{center}
\caption{Modulus couplings to brane gauge bosons in the large-hierarchy geometric regime (using $p=1$ as for the benchmark parameters described in the text).}
\end{table}

\subsubsection*{Branes at singularities}

We next focus on the case where the SM cycle has zero size, $\tau_4\to 0$, with the SM built using fractional $D$-branes located at the singularity. Canonical normalisation and the mass spectrum are also computed in detail elsewhere in this case for the small hierarchy so we simply state the main results, extending them also to the case of large hierarchies.

The effective field theory at the singularity admits a K\"{a}hler potential that can be expanded as \cite{quiver}:
\be
 K=-2\ln\left(\vo'+\frac{s^{3/2}\xi}{2}\right)+\lambda \frac{\tau_4^2}{\vo'}-\ln\left(2 s\right),
\label{Kquivero}
\ee
with $\vo'=\alpha\left(\sqrt{\tau_1}\tau_2-\gamma_3\tau_3^{3/2}\right)$. In (\ref{Kquivero}) we leave the dependence on the real part of the axio-dilaton $s=$Re$(S)$ explicit, even though this modulus is flux-stabilised (in the perturbative regime, $\langle s \rangle=g_s^{-1}>1$) at tree level. We do so because in this case the SM gauge coupling is given by $s$ plus a flux-dependent correction in $\tau_4$: $4\pi g^{-2}=s+h(F)\tau_4$, and so to work out the coupling of the moduli to the SM gauge bosons, we must derive their mixing with both $s$ and $\tau_4$.

The particular form of the K\"{a}hler potential (\ref{Kquivero}) and $\langle\tau_4\rangle=0$ imply that at leading order there is no mixing between $\tau_4$ and the other moduli, leading to the following canonical normalisation around the minimum \cite{CM}:
\begin{eqnarray}
\delta \tau_1 &\simeq& \sum_{i=1}^2 \omega_{1i}
 \, \delta \phi_i + \frac{\omega_{13}}{\vo^{1/2}} \,
 \delta \phi_3 + \frac{\omega_{1s}}{\vo^{1/2}} \,
 \delta \phi_s \approx
 \sum_{i=1}^2 \omega_{1i} \, \delta \phi_i, \\
 \delta \tau_2 &\simeq& \sum_{i=1}^2 {\omega_{2i}}{\vo}
 \, \delta \phi_i + {\omega_{23}}{\vo^{1/2}} \,
 \delta \phi_3 +{\omega_{2s}}{\vo^{1/2}} \,
 \delta \phi_s \approx
 \sum_{i=1}^2 {\omega_{2i}}{\vo} \, \delta \phi_i, \label{cn2quiverNew} \\
 \delta \tau_3 &\simeq& \frac{\omega_{31}}{\vo^{n}} \,\delta \phi_1
 +\omega_{32} \, \delta \phi_2 + {\omega_{33}}{\vo^{1/2}} \, \delta \phi_{3}
 +\frac{\omega_{3s}}{\vo^{1/2}} \, \delta\phi_s
 \approx {\omega_{33}}{\vo^{1/2}} \, \delta \phi_3, \\
 \delta \tau_4 &\simeq& {\omega_{44}}{\vo^{1/2}} \, \delta \phi_4, \label{cn4quiverNew} \\
 \delta s &\simeq& \frac{\omega_{s1}}{\vo^{1/2+n}} \, \delta \phi_1
 + \frac{\omega_{s2}}{\vo^{1/2}} \, \delta \phi_2
 +\frac{\omega_{s3}}{\vo} \, \delta \phi_3
 + \omega_{ss} \, \delta\phi_s \approx \omega_{ss} \delta\phi_s,
\end{eqnarray}
where $n = \frac13$ in the case of a small hierarchy, while $n=p$ in the large-hierarchy case (with $p \simeq 1$ using our numerical benchmark values). The spectrum of modulus masses in this case is
\be
 m_1 \sim \sqrt{\frac{g_s}{4\pi}}\frac{M_p}{\vo^{(3+n)/2}} , \text{ \ }
 m_2 \sim \sqrt{\frac{g_s}{4\pi}}\frac{M_p}{\vo^{3/2}} , \text{ \ }
 m_3 \sim m_s \sim \sqrt{\frac{g_s}{4\pi}}\frac{M_p}{\vo} \text{ \ and \ }
 m_4 \sim \sqrt{\frac{g_s}{4\pi}}\frac{M_p}{\vo^{1/2}} \,,
\label{MASSES}
\ee
since both $s$ and $\tau_4$ are fixed at order $\vo^{-2}$ but $K^{-1}_{s\bar s}\sim\mc{O}(1)$ while $K^{-1}_{4\bar 4}\sim\mc{O}(\vo)$.

\medskip\noindent {\em Couplings:}

\medskip\noindent
The volume scaling of the $(\zeta_i/M_p) \, \delta \phi \, F_{\mu\nu}^{(i)} F^{\mu\nu}_{(i)}$ couplings to brane gauge bosons, with $i=1,2,3,4$ is summarised in Tables 4 and 5 \cite{CM}, which again reveal a rich pattern of couplings varying from weak-interaction strength ($\sim \vo^{1/2}/M_p$), gravitational strength ($\sim 1/M_p$) and much weaker than gravitational strength ($\sim \vo^{-k}/M_p$ with $k>0$).

\begin{table}[ht]
\begin{center}
\begin{tabular}{c|ccccc}
  & $\delta\phi_1$ & $\delta\phi_2$
  & $\delta\phi_3$ & $\delta\phi_4$ & $\delta\phi_s$ \\
  \hline\hline
  \\ & & & & \vspace{-0.9cm}\\
  $\zeta_{1}, \zeta_{2}$
  & $1$
  & $1$
  & $\vo^{-1/2}$
  & 0
  & $1$ \\
  \\ & & & & \vspace{-0.9cm}\\
  $\zeta_{4}$
  & $\vo^{-5/6}$
  & $\vo^{-1/2}$
  & $\vo^{-1}$
  & $\vo^{1/2}$
  & $1$
\end{tabular}
\end{center}
\caption{Modulus couplings to brane gauge bosons in the small-hierarchy singular regime.}
\end{table}

\begin{table}[ht]
\begin{center}
\begin{tabular}{c|ccccc}
  & $\delta\phi_1$ & $\delta\phi_2$
  & $\delta\phi_3$ & $\delta\phi_4$ & $\delta\phi_s$ \\
  \hline\hline
  \\ & & & & \vspace{-0.9cm}\\
  $\zeta_{3}$
  & ${\vo^{-p}}$
  & $1$
  & $\vo^{1/2}$
  & 0
  & $1$ \\
  \\ & & & & \vspace{-0.9cm}\\
  $\zeta_{4}$
  & $\vo^{-1/2-p}$
  & $\vo^{-1/2}$
  & $\vo^{-1}$
  & $\vo^{1/2}$
  & $1$
\end{tabular}
\end{center}
\caption{Modulus couplings to brane gauge bosons in the large-hierarchy singular regime (with $p=1$ for our choice of benchmark parameters).}
\end{table}

Even in this case the volume scaling of the modulus normalisation and couplings can be understood from a geometrical point of view. Focusing for example on $\delta\phi_3$, we notice that its coupling to $F_{\mu \nu}^{(3)} F^{\mu \nu}_{(3)}$ is stronger than the coupling to $F_{\mu \nu}^{(1,2)} F^{\mu \nu}_{(1,2)}$ which, in turn, is stronger than the coupling to the SM gauge bosons $F_{\mu \nu}^{(4)} F^{\mu \nu}_{(4)}$. This different behaviour reflects the fact that $\tau_3$ resolves a point-like singularity which has a definite location within the Calabi-Yau, together with the sequestering of the SM at the $\tau_4$-singularity.

\subsection{Modulus spectrum and couplings: corrections}

We now estimate the size of various correction to the above modulus masses, to study their robustness against higher loops. As is true for generic large-volume models, the majority of loop corrections that one might naively expect to dominate do not do so because they are suppressed by the accumulated powers of $1/\vo$ appearing in the masses and couplings. As argued in ref.~\cite{uber}, such suppressions are a general consequence of having the gravity scale very small compared with the Planck scale.

However, there are two kinds of loop contributions that are particularly dangerous for the models of interest here, and we now estimate their size. The two kinds of corrections are: mixings amongst the moduli that are induced by loop contributions to gauge kinetic terms, as described in ref.~\cite{joeredef}; and corrections due to loops of heavy particles on the supersymmetry-breaking SM brane, as described in ref.~\cite{uber}.

\subsubsection*{Corrections to gauge kinetic terms}

At the one-loop level it can happen that the physical modulus is not as simply related to the holomorphic modulus as it is classically. In particular, threshold corrections to the gauge kinetic terms can introduce large logarithms into the definitions of the gauge couplings, of the form $\ln(M'/M)$ where $M'$ and $M$ are the masses of two kinds of massive states that have been integrated out. But because we have seen that different states can have masses that depend on different powers of $\vo$, for large-volume compactifications such logarithms need not be either small or holomorphic.

In particular, it can happen that the physical modulus, $\tau_4$, controlling the blow-up cycle for a singularity becomes related to the holomorphic modulus, $\tau_4'$, through the relation
\be \label{holoredef}
 \tau_4 = \tau_4' - \kappa\ln\vo \,,
\ee
where $\kappa$ is an $\mc{O}(1)$ constant. This redefinition is also required in order to have an effective supergravity description that is consistent with the general Kaplunovsky-Louis formula \cite{CL} for the running of the gauge coupling \cite{1loopf}.

Notice in particular that the holomorphic modulus need not then vanish in the singular limit where the volume of the blow-up cycle shrinks to zero; {\em i.e.} when $\langle\tau_4\rangle=0$. Hence in this case the holomorphic SM modulus takes a nonzero VEV at the singularity, and when $\vo$ is large this VEV can be comparable to that of a generic blow-up mode within the geometric regime. Although this is not a significant change for branes that are already in the geometric regime, having $\langle \tau_4' \rangle$ nonzero can (but need not) significantly change the predictions for branes localised at singularities. Detailed studies \cite{joeredef} show that this kind of correction really does arise for combinations of $D3s/D7s$ located at orbifold singularities as well as the phenomenologically less interesting case of $D3s$ at orientifold singularities, but does {\em not} arise if there are only $D3s$ situated at orbifold points.

To see why the redefinition can change predictions for masses and couplings, recall that it is the holomorphic field that transforms in the standard way under 4D supersymmetry and so appears in the standard 4D supergravity action. Since a non-holomorphic redefinition like eq.~\pref{holoredef} can change the form of the kinetic terms, it also changes the transformations required to achieve canonical normalisation. In particular, the K\"{a}hler potential (\ref{Kquivero}) and the gauge coupling in this case get modified to
\begin{gather}
 K = -2 \ln\left( \vo' + \frac{s^{3/2}\xi}{2} \right) + \frac{\lambda \left( \tau_4' - \kappa \ln\vo' \right)^2}{\vo'} - \ln\left( 2 s \right), \label{NewKquiver} \\
 4 \pi g^{-2} = s + h(F) \left( \tau_4' - \kappa \ln\vo' \right), \text{ \ with \ } \vo' = \alpha \left( \sqrt{\tau_1} \tau_2 - \gamma_3 \tau_3^{3/2} \right).
\label{KquiverNew}
\end{gather}

The new K\"{a}hler potential (\ref{NewKquiver}) yields additional contributions to the kinetic terms of the 4D fields, of the form
\begin{gather}
 \mathcal{L}_{\rm kin}^{\rm new} = - \frac{\kappa\lambda}{\langle\vo\rangle}
 \left[ \frac{\partial_{\mu}(\delta\tau_1)}{2\langle\tau_1\rangle}
 + \frac{\partial_{\mu}(\delta\tau_2)}{\langle\tau_2\rangle} \right]
 \partial^{\mu}(\delta\tau_4') + \frac{3\alpha\kappa\lambda\gamma_3 \sqrt{\langle\tau_3\rangle}}{2\langle\vo\rangle^2}
 \,\partial_{\mu}(\delta\tau_3) \, \partial^{\mu}(\delta\tau_4') \notag \\
 =  \frac{\kappa\lambda}{\langle\vo\rangle^2} \left[ - \partial_{\mu}(\delta\vo)
 +\frac32 \, \alpha\gamma_3\sqrt{\langle\tau_3\rangle}
 \,\partial_{\mu}(\delta\tau_3) \right] \partial^{\mu}(\delta\tau_4')
\label{LkinquiverNew}
\end{gather}
which give rise to a non-vanishing mixing between $\tau_4'$ and all the other moduli but the dilaton. Notice that this mixing is absent if there is no 1-loop redefinition, i.e. $\kappa=0$. Therefore the canonical normalisation (\ref{cn4quiverNew}) changes from $\delta \tau_4 \simeq {\omega_{44}}{\vo^{1/2}} \, \delta \phi_4$ to
\be
 \delta \tau_4' \simeq \frac{\omega_{41}}{\vo^{n}} \,\delta \phi_1
 +\omega_{42} \, \delta \phi_2 + \frac{\omega_{43}}{\vo^{1/2}} \, \delta \phi_3
 +{\omega_{44}}{\vo^{1/2}} \,\delta \phi_4 \,,
 \label{cn4quiverRedef}
\ee
which adds a mixing with $\delta \phi_1$, $\delta \phi_2$ and $\delta \phi_3$ not present in eq.~\pref{cn4quiverNew}.
Inspection of eq.~(\ref{CANnorm3}) shows these new mixings scale with $\vo$ in the same way as for the geometric regime, but with the difference that the mixing between $\delta \phi_4$ and $\delta \phi_3$ is suppressed (because the two blow-up cycles do not intersect and so do not experience the $\mc{O}(1)$ mixing).

Chasing the effects of this change in canonical normalisation (\ref{cn4quiverRedef}) and the gauge kinetic function (\ref{KquiverNew}) through to the modulus/gauge-field couplings yields the couplings to gauge bosons shown in Tables 6 and 7.

\begin{table}[ht]
\begin{center}
\begin{tabular}{c|ccccc}
  & $\delta\phi_1$ & $\delta\phi_2$
  & $\delta\phi_3$ & $\delta\phi_4$ & $\delta\phi_s$ \\
  \hline\hline
  \\ & & & & \vspace{-0.9cm}\\
  $\zeta_{1,2}$
  & $1$
  & $1$
  & $\vo^{-1/2}$
  & $\vo^{-1/2}$
  & $1$ \\
  \\ & & & & \vspace{-0.9cm}\\
  $\zeta_4$
  & $\vo^{-1/3}$
  & $1$
  & $\vo^{-1/2}$
  & $\vo^{1/2}$
  & $1$
\end{tabular}
\end{center}
\caption{Modulus couplings to brane gauge bosons for the small-hierarchy singular regime, including the 1-loop modulus redefinition.}
\end{table}

\begin{table}[ht]
\begin{center}
\begin{tabular}{c|ccccc}
  & $\delta\phi_1$ & $\delta\phi_2$
  & $\delta\phi_3$ & $\delta\phi_4$ & $\delta\phi_s$ \\
  \hline\hline
  \\ & & & & \vspace{-0.9cm}\\
  $\zeta_{3}$
  & $\vo^{-p}$
  & $1$
  & $\vo^{1/2}$
  & ${\vo^{-1/2}}$
  & $1$ \\
  \\ & & & & \vspace{-0.9cm}\\
  $\zeta_{4}$
  & $\vo^{-p}$
  & $1$
  & ${\vo^{-1/2}}$
  & ${\vo^{1/2}}$
  & $1$
\end{tabular}
\end{center}
\caption{Modulus couplings to brane gauge bosons for the large-hierarchy singular regime including 1-loop modulus redefinition (with $p=1$ for the benchmark parameters used in the text).}
\end{table}

As expected, the 1-loop redefinition makes the modulus couplings scale the same way with $\vo$ as do those of the geometric regime, as summarised in Tables 2 and 3 (with the difference that now $\tau_3$ does not intersect $\tau_4$). Here $\vo$ is given by expressions (\ref{volgeom}) and (\ref{volsing}) in the geometric and singular cases, respectively. Thus the coupling of $\delta\phi_3$ to the SM gauge bosons living on $\tau_4$ is more $\vo$-suppressed than the coupling of $\delta\phi_4$ to the same visible degrees of freedom due to the geometric separation of the two point-like singularities resolved by these two different blow-up modes.

\subsubsection*{Corrections to the scalar potential}

Loop corrections to low-energy scalar potentials are notoriously sensitive to the details of the theory's UV sector, and so must be examined carefully for any calculation that predicts small scalar masses. For instance, the one-loop corrections to the low-energy scalar potential in four dimensions has the Coleman-Weinberg form \cite{ColeWbg}
\be \label{CWPot}
 \delta V_{CW}^{1-loop}(\varphi) \propto \hbox{STr} \left\{ M^4(\varphi) \ln \left[ \frac{M^2(\varphi)}{\mu^2} \right] + c_1 M^2 + c_0 \right\} \,,
\ee
where STr denotes the usual spin- and statistics-weighted sum over heavy degrees of freedom circulating in the loop. Here $M(\varphi)$ denotes the renormalised\footnote{For reasons discussed elsewhere \cite{uber,uses} we formulate UV sensitivity in terms of large physical masses rather than cutoffs.} mass matrix of the particles circulating in the loop --- regarded as a function of the low-energy scalar fields, $\varphi$; the constants $c_1$ and $c_0$ and the floating renormalisation point, $\mu$,  depend on the precise renormalisation scheme used.

The bad news is that eq.~\pref{CWPot} involves positive powers of $M^2$ and so can depend sensitively on the UV spectrum. The good news is that eq.~\pref{CWPot} holds only in 4 dimensions, and so in higher-dimensional theories the largest value of $M$ that can appear is the KK scale above which the UV theory becomes higher dimensional. Of course the low-energy potential might still be sensitive to the contributions of higher-energy modes, but this sensitivity must be computed in the full higher dimensional theory (where additional symmetries, like higher-dimensional general covariance can play a r\^ole). In particular, contributions from states at the string scale are described by the usual local, higher-derivative terms that capture the well-known $\alpha'$ corrections. Since for LARGE-volume models these are already included in what we are calling the `leading-order' corrections, they do not destabilise any of the conclusions found above.

It is a remarkable feature of theories with low gravity scales --- including the large-$\vo$ models of interest here --- that loop corrections to the low-energy scalar potential are smaller than would have been indicated by a 4D expression like eq.~\pref{CWPot}. As is argued in ref.~\cite{uber}, this happens both because the KK scale that must be used in eq.~\pref{CWPot} is so low, and because higher-dimensional symmetries constrain the kinds of contributions that can arise from states much more massive than the KK scale. Of course, loop estimates are much harder for the very anisotropic geometries considered here, having many scales between $M_s$ and $M_\KK^{6D}$. We have sought higher-loop, extra-dimensional contributions that can use this complication, and we now describe the largest we have found.

The sector we find to be the most dangerous in loops consists of those open-string states localised on the SM brane itself. These are dangerous because of precisely the same features that were required earlier in this section for successful phenomenology: ($i$) they must badly break supersymmetry; and ($ii$) they must reside on a very small cycle. In particular we know there is a non-supersymmetric 4D sector localised on the SM brane up to masses of order $\alpha^{1/4}_\SM M_s \simeq 0.3 M_s$. Since these are effectively 4D up to these scales, eq.~\pref{CWPot} applies and predicts contributions of generic size
\be \label{1loopVsinV}
 \delta V_{CW}^{1-loop}(\varphi) \simeq M_s^4 + m_{3/2}^2 M_s^2 + \cdots
 \simeq \frac{M_p^4}{\vo^2}+ \frac{M_p^4}{\vo^3} + \cdots \,.
\ee

It is useful to compare this estimate with the size of the leading stabilisation contributions to the low-energy potential described in \S2. There we found the leading terms are $V_{\rm lead} \sim M_p^4/\vo^3$ but depend only on the moduli $\vo$ and $\tau_3$. The masses of the rest of the K\"ahler moduli that do not appear in $V_{\rm lead}$ then come from subdominant terms, of order $\delta V \sim M_p^4/\vo^k$, with $k>3$. Because these are subleading in $1/\vo$
with respect to the terms in eq.~\pref{1loopVsinV}, loop effects on non-supersymmetric localised branes cannot be negligible, and besides yielding potentially large corrections to the modulus masses might also destabilise the vacuum itself. This may yet prove to be a feature rather than a bug, since it may be the source of the unknown physics responsible for lifting the present-day vacuum energy to near zero. Concrete attempts to use brane back-reaction to address the cosmological constant problem can be found in \cite{SLED,SLEDrev,5Dcc}.

The lesson to be drawn from this observation is that brane loops and brane back-reaction can become important for understanding the full dynamics of the vacuum when non-supersymmetric branes appear in large-volume models. Although it remains an unsolved problem as to how this dynamics works in string theory, the effects of on-brane loops \cite{6DLoops} and back-reaction \cite{6DMatching, 6DBraneBR} have been studied for non-supersymmetric, codimension-2 branes in simpler 6D extra-dimensional models. These simpler systems resemble their 10D cousins in that the back-reaction is also competitive with bulk physics in stabilising the extra dimensions, yet doesn't destroy the presence of large-volume solutions. Intriguingly they can also allow on-brane curvatures to be parametrically smaller than naive estimates based on the brane tensions would suggest \cite{6DBraneBR}.

For the present purposes we assume the SM back-reaction not to destroy the broad properties of the flux compactification described to this point, and ask how these radiative corrections change the masses and couplings of the moduli. These are generically of order
\be
 \delta m \simeq \frac{\zeta \, M_s^2}{M_p} \simeq \frac{\zeta \, M_p}{\vo}  \,,
\ee
where the modulus-brane coupling $\zeta$ is as given in the earlier Tables, and as before we take $M_s$ to be the UV mass scale on a non-supersymmetric brane. Notice that when $\zeta$ is suppressed by inverse powers of $\vo$ this correction can be smaller than the generic modulus mass, $\delta m \ll  M_p/\vo$.

\begin{table}[ht]
\begin{center}
\begin{tabular}{c|cccc}
  & $m_1/M_p$ & $m_2/M_p$ & $m_3/M_p$ & $m_4/M_p$ \\
  \hline\hline
  \\ & & & & \vspace{-0.9cm}\\
  leading
  & $\vo^{-(3+n)/2}$ & $\vo^{-3/2}$
  & $\vo^{-1}$ & $\vo^{-m/2}$ \\
  \\ & & & & \vspace{-0.9cm}\\
  loop
  & $\vo^{-s_n}$ & $\vo^{-1}$
  & $\vo^{-1/2}$ & $\vo^{-1/2}$
\end{tabular}
\end{center}
\caption{Leading and loop-corrected masses for K\"ahler moduli in the geometric regime. $n = \frac13$ for the small-hierarchy case, and $n=1$ for the benchmark large-hierarchy example. $s_n = \frac12(3+n)$ when $n \ge 1$ and $s_n = 1+n$ if $n \le 1$. The parameter $m = 1$ when
$D$-terms stabilise the SM brane and $m = 2$ if this is done using $g_s$ corrections.}
\end{table}

\begin{table}[ht]
\begin{center}
\begin{tabular}{c|cccc}
  & $m_1/M_p$ & $m_2/M_p$ & $m_3/M_p$ & $m_4/M_p$ \\
  \hline\hline
  \\ & & & & \vspace{-0.9cm}\\
  leading
  & $\vo^{-(3+n)/2}$ & $\vo^{-3/2}$
  & $\vo^{-1}$ & $\vo^{-1/2}$ \\
  \\ & & & & \vspace{-0.9cm}\\
  loop (pot only)
  & $\vo^{-(3+n)/2}$ & $\vo^{-3/2}$
  & $\vo^{-1}$ & $\vo^{-1/2}$ \\
  loop (pot and mix)
  & $\vo^{-s_n}$ & $\vo^{-1}$
  & $\vo^{-1}$ & $\vo^{-1/2}$
\end{tabular}
\end{center}
\caption{Leading and loop-corrected masses for K\"ahler moduli in the singular regime. The first loop estimate excludes changes due to loop-generated mixing among moduli (as is appropriate for some models), while the second includes this mixing (as appropriate for other models - see text for a description of which is which). $n = \frac13$ for the small-hierarchy case, and $n=1$ for the benchmark large-hierarchy example. $s_n = \frac12(3+n)$ when $n \ge 1$ and $s_n = 1+n$ if $n \le 1$. }
\end{table}

Tables 8 and 9 then show how this loop estimate changes the predictions for modulus masses for both the geometric and singular regimes, with $n = \frac13$ appropriate for small hierarchies and $n =1$ is the value used in the large-hierarchy benchmark given in \S2. (Only the $n = \frac13$ geometrical case is considered in ref.~\cite{uber}, and agrees with the values shown here.) These tables, when combined with the earlier tables of coupling strengths, paint an interestingly complex picture. In it the state $\delta \phi_4$ is revealed to be pulled up in mass to join the highest-mass KK states, as appropriate for the modulus of a localised cycle. Its couplings to fields on this cycle are of order $\sim 1/M_{10D}$ appropriate to higher-dimensional gravity, making them weak-interaction strength when $M_s$ is at TeV scales.

A similar thing happens in the geometric regime to $\delta \phi_3$, which is associated with the other localised cycle. In the singular regime this state instead remains in the same mass range, $M_p/\vo$, that is generic for moduli. Remarkably, this modulus couples to SM brane fields with much weaker than gravitational strength.

The state $\delta \phi_2$ --- which dominantly corresponds to the volume modulus --- is also generically lifted by loops from its initially smaller value, but in this case only as high as $M_p/\vo$. When all of the dimensions have the same size, their common KK scale is $M_p/\vo^{2/3}$ and so these moduli remain well within the low-energy 4D description. But in the large-hierarchy scenario the generic moduli are close in mass to the lightest of the 6D KK states and can become lost into mixings with more generic KK modes, potentially losing their 4D interpretation.

The fibre modulus, $\delta \phi_1$, is more unusual for several reasons. First, it is the simplest state that is orthogonal to the volume modulus, for which the potential arises at sub-dominant order in $1/\vo$. This is why the leading term in its mass is so small.\footnote{It is this small leading mass that motivates using this state as the inflaton in `Fibre Inflation' models \cite{fiberinfl}
and as the curvaton in models with non-standard primordial fluctuations and large non-gaussianities \cite{BCGQTZ}.} In LARGE-volume vacua such a modulus
does not arise unless there are at least three K\"ahler moduli, and so they are not present in the very simple compactifications most often explored. But it is also unusual because although loops lift it from its leading, extremely small, mass, its small coupling to the SM brane ensures they do not lift it very far \cite{uber}. Although its couplings to other states -- like bulk KK modes for instance -- need not be equally suppressed, loops of these appear to remain suppressed by the same general covariance and supersymmetry arguments that generally apply for states deep in the extra-dimensional regime.

\subsection{Bulk Kaluza-Klein modes}

For later convenience we close this section by noting a few properties of generic, non-modulus, bulk KK modes, such as for the metric $h_{\ssM \ssN}$, Kalb-Ramond fields, $B_{\ssM\ssN}$, the axio-dilaton and so on. For TeV-scale strings there is always a great abundance of these modes, with 10D kinematics extending down to energies of order $M_\KK^{10D}$, and 6D kinematics continuing down to $M_\KK^{6D}$. A small number of states --- including the 4D graviton and moduli --- survive below this scale into the 4D theory.

Although we've seen that the moduli can couple with weaker than gravitational strength, this is typically to do with having a small overlap with a localised cycle and should not be true for generic higher KK modes that are free to move throughout the bulk (and are not localised in warped throats, say \cite{warping,GA}). KK modes with short wavelengths that are free to move about the geometry should couple with gravitational strength, $1/M_p$, just as they do in simpler geometries like spheres or tori.

An important difference compared with tori and spheres is the absence of continuous isometries on compact Calabi-Yau spaces. Unless broken by other fields isometries show up as unbroken symmetries in the low-energy 4D theory, under which some KK states are charged. This makes the lightest charged KK mode stable, with important phenomenological consequences. Since Calabi-Yau spaces have none, their KK modes are not protected in this way. It is nevertheless possible to have isometries for {\em submanifolds} of compact Calabi-Yau spaces, and if so states localised near these submanifolds can be charged under approximate symmetries that make them very long-lived. This makes it important to identify such submanifolds for candidate Calabi-Yau vacua, and see whether KK modes actually do localise near them.

\section{Phenomenological issues}

With mass scales and spectra in hand, it is possible to address --- at least in a preliminary way --- some of the phenomenological features to be expected of these models. Of particular interest is the way knowledge of the UV completion provides more information about the low-energy limit than is generic to a garden-variety phenomenological model with a low gravity scale. What we find must contain the generic predictions of supersymmetric large extra dimensions \cite{MSLED}, but extends these by providing the more detailed predictions for the low-energy spectrum and couplings that the UV completion makes possible. Both the low-energy bulk supersymmetry and the new states make the predictions differ significantly from those of minimal `ADD' models \cite{ADDpheno,ADDphenoCollider}, for which gravity is the only field that propagates in the bulk.

\subsubsection*{Scales}

We first summarise the predicted mass scales for the main alternatives. For convenience these are tabulated in Table 10 for both the geometric and singular regimes, including loop corrections to the various masses (with and without modulus mixing in the singular case). The numerical values use the benchmarks defined in \S2.

\begin{table}[ht]
\begin{center}
\begin{tabular}{c|ccc|ccc}
  hierarchy && small &&& large & \\
  regime & geo & sing (w mix) & sing & geo & sing (w mix) & sing \\
  \hline\hline
  \\ & &  & &  & & \vspace{-0.9cm}\\
  $M_s$
  & 1 TeV & 1 TeV & 1 TeV & 3 TeV & 3 TeV & 3 TeV \\
  \\ & &  & &  & & \vspace{-0.9cm}\\
  $M_{6D}$
  & 2000 TeV & 2000 TeV & 2000 TeV & 10 TeV & 10 TeV & 10 TeV \\
  $M_{10D}$
  & 2 TeV & 2 TeV & 2 TeV & 4 TeV & 4 TeV & 4 TeV \\
  $M_\KK^{c}$
  & 0.5 TeV & 0.5 TeV & 0.5 TeV & 1 TeV & 1 TeV & 1 TeV \\
  $M_\KK^{10D}$
  & 50 MeV & 50 MeV & 50 MeV & 1 TeV & 1 TeV & 1 TeV \\
  $M_\KK^{6D}$
  & 0.3 MeV & 0.3 MeV & 0.3 MeV & 0.01 eV & 0.01 eV & 0.01 eV \\
  $m_{3/2}$
  & $10^{-3}$ eV & $10^{-3}$ eV & $10^{-3}$ eV & $10^{-3}$ eV & $10^{-3}$ eV & $10^{-3}$ eV \\
  $m_{moduli}$
  & 0.01 eV & 0.01 eV & 0.01 eV & 0.01 eV & 0.01 eV & 0.01 eV  \\
  $m_{2}$
  & 0.01 eV & 0.01 eV & $10^{-17}$ eV & 0.01 eV & 0.01 eV & $10^{-17}$ eV \\
  $m_1$
  & $10^{-12}$ eV & $10^{-12}$ eV & $10^{-22}$ eV & $10^{-32}$ eV & $10^{-32}$ eV & $10^{-32}$ eV
\end{tabular}
\end{center}
\caption{Numerical (loop-corrected) spectrum for the geometric and singular regimes. This uses $n = \frac13$ for small-hierarchies and $n=1$ for the benchmark large-hierarchy. For the masses of the two light moduli, $m_1$ and $m_2$, the powers of $n$ numerically become $M_p/\vo \sim 0.01$ eV, $M_p/\vo^{4/3} \sim 10^{-12}$ eV, $M_p/\vo^{3/2} \sim 10^{-17}$ eV, $M_p/\vo^{5/3} \sim 10^{-22}$ eV and $M_p/\vo^2 \sim 10^{-32}$ eV.}
\end{table}

The main difference revealed by the table is that between large and small hierarchies, since this dramatically changes the scale at which the lightest KK state arises. This difference is similar to the usual difference between phenomenological models having two or more large extra dimensions.

The table also shows that for both large- and small-hierarchy geometries the spectrum is similar when the SM is wrapped on a geometric and singular cycle, provided that there is loop mixing among the moduli (as is the case for most systems of practical interest, such as those involving $D7$s at orbifold points). It is only when this mixing is absent that the spectrum differs for a singular cycle, and the most important difference is a suppression of the mass of the volume modulus, $m_2$.

\begin{table}[ht]
\begin{center}
\begin{tabular}{c|ccccc}
  $m$
  & $10^{-2}$ eV & $10^{-12}$ eV & $10^{-17}$ eV
  & $10^{-22}$ eV & $10^{-32}$ eV \\
  \\ &  & &  & & \vspace{-0.9cm}\\
  $m^{-1}$
  & 10 $\mu$m & 100 km & 0.1 AU & 0.1 ly & 300 Mpc \end{tabular}
\end{center}
\caption{Ranges relevant to force tests for various choices of modulus masses.}
\end{table}

Finally, the table shows that both the volume- and the fibre-modulus masses, $m_2$ and $m_1$, can be remarkably light even once loops are included. (Table 11 gives the ranges in more conventional units over which particles this light can interact coherently, and so over which can mediate new forces.) Such small masses are stable against loops because of the very weak couplings between these particles and the supersymmetry-breaking SM sector, and we argue below that these weak couplings also suppress their contributions to  macroscopic tests of gravity (to which they can contribute because of their small masses). We discuss the implications of, and uncertainty in, these masses and couplings in more detail below.

\subsection{SLED-related constraints}

As might be expected for a theory with so many exotic light states, models of this type are subject to a variety of stringent constraints. These come in two broad classes: those that are generic to having supersymmetric large dimensions, and those that arise because of the presence of specific types of new light fields. We briefly discuss each in turn, starting here with the most robust and generic consequences: those following just from the existence of supersymmetric large dimensions.

\subsubsection*{Missing energy and KK exchange}

The most robust signature to occur in systems with large dimensions is energy loss into the extra dimensions, since this assumes nothing about the branching rate for KK modes to produce visible SM particles. Signals coming from the virtual exchange of extra-dimensional particles are also possible, but are more model-dependent to interpret since they assume the absence of exotic decay processes \cite{MSLED} and since exchange also competes with unknown direct brane-localised contact interactions that need not involve the extra dimensions at all \cite{KKXchange}. Because large dimensions were initially proposed \cite{ADD} as alternatives to supersymmetry, the study of this loss rate is usually aimed specifically at the radiation of extra-dimensional gravitons \cite{ADDphenoCollider}, since this has the advantage that the graviton couplings are relatively model-independent.

Emission cross sections can be sizable because of the enormous phase space of states that can be emitted; even though each KK graviton mode couples with 4D gravitational strength, $\sigma_n \propto 1/M_p^2$, the sum over all modes converts this small coupling to higher-dimensional gravitational strength. For $d$ extra dimensions $\sigma \sim \sum_n \sigma_n \propto (V_d E^d)/M_p^2 \propto E^d/M_{\ssD}^{2+d}$, where $M_{\ssD}^{2+d} = (8 \pi G_\ssD)^{-1}$ is the reduced Planck scale in $D= 4+d$ dimensions. For $M_{\ssD} \sim 1$ TeV this leads to weak-interaction production rates.

Two consequences follow from the strong growth with energy of these cross sections,
\be
 \sigma \propto \frac{1}{M_{\ssD}^2} \left( \frac{E}{M_{\ssD}} \right)^d \,.
\ee
First, such strong growth would eventually violate the unitarity bound once $E \sim M_{\ssD}$, indicating that a fuller string calculation is required at higher energies, where the emission and exchange of string states is no longer negligible \cite{HEGrav}. Second, it shows that it is the highest energy KK states that dominate in the cross section, and since these also have the shortest wavelength their properties (and the cross section) is largely insensitive to the details of the higher-dimensional geometry \cite{Leblond}. Consequently cross sections at high energies --- such as for processes at colliders --- that are computed using simple toroidal models for the extra dimensions are also likely to capture those for more complicated Calabi-Yau extra-dimensional geometries.

The absence of this observed energy-loss signal in an experiment at a given energy $E$ can be quoted as an upper bound on the extra-dimensional gravity scale, $M_{\ssD}$. Because $d$ controls the power of the small ratio $E/M_\ssD$ in the cross section, the bound on $M_\ssD$ obtained from a fixed $E$ and upper limit on $\sigma(E)$ weakens with growing $d$. Searches for graviton emission at the Tevatron \cite{ColliderBoundsTev} place limits of order $M_\ssD \gsim 1$ TeV for $d = 2$ (so $D=6$) and $M_\ssD \gsim 0.8$ TeV for $d = 6$ (and so $D=10$).

Since supersymmetry introduces many more states into the bulk than just the graviton, there are potentially many more channels for energy loss when the large extra dimensions are supersymmetric \cite{MSLED,SLEDpheno,SLEDscalar}. This means that the relation assumed between $\sigma(E)$ and $M_\ssD$ differs in the supersymmetric case from vanilla ADD models where only the graviton appears. To the extent that all of these new states also couple to the SM brane with a strength similar to the bulk graviton, on dimensional grounds their production cross section scales the same way with $E$, leading to an estimate $\sigma_{\scriptscriptstyle SLED} \sim N \sigma_{\scriptscriptstyle LED}$ where $N$ is an estimate of the number of additional states present in the bulk. But because $\sigma \propto M_{6D}^{-4}$ when $d=2$, this means that it is really $M_{6D}/N^{1/4}$ that is constrained rather than $M_\ssD$ when upper limits on $\sigma(E)$ are compared with calculations assuming only graviton emission. Luckily, this represents a factor of $\sim 3$ even if $N \sim 100$ \cite{MSLED}.

An important assumption in this estimate is that the new field content also couples to branes with gravitational strength. But this need not be true, particularly given that the SM brane must badly break supersymmetry. Better yet, in 6D there are some kinds of bulk fields for which dimensionless couplings are possible, such as a coupling $\int \exd^4x \, H^\dagger H \, \phi$ where $H$ is the usual Higgs boson and $\phi$ is an extra-dimensional scalar. In this case the emission cross section need not grow as a power of $E/M_{6D}$, and so can extend the reach of extra-dimensional searches \cite{SLEDscalar}.

More recent limits are also available from the LHC, however to date these rely on exchange processes for which the produced extra-dimensional states are assumed to decay into visible particles (and so are slightly more model-dependent). These give slightly larger bounds, $M_\ssD \gsim 4$ TeV \cite{ColliderBoundsLHC,ColliderBoundsLHCth}.

\subsubsection*{Supersymmetric phenomenology}

The models examined here share another robust consequence of supersymmetric extra dimensions: the absence of MSSM superpartners for each of the known SM particles, despite $m_{3/2}$ being so low. This occurs because having $M_s$ as low as TeV scales implies the SM must reside on a non-supersymmetric brane. As a result supersymmetry is nonlinearly realised: applying a supersymmetry transformation to a particle like the electron gives the electron plus a goldstino and {\em not} a selectron \cite{MSLED}. This means that the spectrum on the SM brane does not include the MSSM, implying the --- so far very successful \cite{LHCSUSY} --- prediction that LHC searches for MSSM states should find none. SLED models are the remarkable counter-example to the assertion that weak-scale supersymmetry requires the MSSM.

\subsubsection*{Astrophysical bounds}

Astrophysical systems provide strong constraints on large extra dimensions due to the new energy-loss channels such dimensions would provide for stellar systems and supernovae \cite{ADDpheno,SNothers,HR}. Perhaps surprisingly, these bounds are even stronger than collider limits despite the much lower energies to which they have access: ambient temperatures set the typical energies as $E \sim T \sim 10$ MeV. Again the limits obtained come in two forms: model-independent constraints on energy loss; and more model-dependent bounds that assume specific branching ratios of KK states into ordinary particles.

In particular, standard calculations of supernova energy loss agree well with SN1987A observations, and if this is interpreted as an upper limit on energy-loss into gravitons it implies $M_{6D} \gsim 9$ TeV for 2 extra dimensions \cite{HR}. Because of the comparatively low energies involved, this bound collapses to $M_{10D} > 10$ GeV if all six extra dimensions are similar in size. Although this latter bound is easily evaded for the small-hierarchy models considered here, they provide a stronger test for the large-hierarchy case since they robustly require $M_{6D} \gsim 10$ TeV.

Much more stringent bounds are also possible if the extra-dimensional KK modes have significant branching fractions into observable particles, like photons or gluons \cite{SNothers,HR}. In this case the absence of a $\gamma$-ray signal in the EGRET satellite implies $M_{6D} \gsim 40$ TeV for the large-hierarchy case (dropping to $M_{10D} \gsim 40$ GeV when all six dimensions are similar in size). Considerations of neutron-star cooling give even stronger limits: $M_{6D} \gsim 700$ TeV (or $M_{10D} \gsim 200$ GeV). Because they are more model-dependent, these much stronger limits only apply under certain assumptions. In particular they are evaded if the KK modes have much more efficient branching fractions into invisible degrees of freedom \cite{ADDpheno} (such as decays onto another `trash' brane) are faster than those into SM particles.

This makes it important to know precisely how KK modes decay in any particular candidate string vacuum. In the models of interest here the rate for generic KK modes decay onto states localised on both of the small cycles, $\tau_3$ and $\tau_\SM = \tau_4$, are likely to be very similar, provided similar numbers of states are present on each into which the decay can take place. However, cascade decays into lower-energy states, either in the bulk or on branes wrapped on the large cycles (if present), are also possible and would be equally invisible. (Although energy-momentum conservation forbids straight bulk decays for simple toroidal dimensions, they can occur for the more complicated extra-dimensional geometries arising here because of their absence of isometries.) The rate for KK decays to a 4D configuration like the SM brane is of order $\Gamma_\SM \sim M_\KK^3/M_p^2$, while decays to  higher-dimensional states effectively couple with higher-dimensional Planck strength, according to the number of dimensions into which the daughter states can move \cite{ADDpheno}. Decays into effective 6D states therefore have a rate $\Gamma_{6D} \sim M_\KK^5/M_{6D}^4$ while those that can decay into 10D states do so with rate $\Gamma_{10D} \sim M_\KK^9/M_{10D}^8$. If all three are possible, they occur with the relative rates
\be
 \Gamma_\SM : \Gamma_{6D} : \Gamma_{10D} \sim \frac{M_\KK^3}{L^2 l^4} : \frac{M_\KK^5}{l^4} : M_\KK^9 \,,
\ee
showing that (all other things being equal) decays into 6D and 10D states dominate those to the SM brane by a factor of $L^2/l^2$ if $M_\KK \sim 1/l$. For states with $M_\KK \sim 10$ MeV using $1/L \sim 0.01$ eV and $1/l \sim 1$ TeV gives $\Gamma_\SM / \Gamma_{6D} \sim 10^{-18}$, while decays to 10D states are not energetically allowed. This shows that decays to 6D states can indeed dominate for the states most relevant to astrophysics.

Ensuring that astrophysical energy-loss bounds are not transgressed is an important constraint on more precise models of the physics on the various branes. The complicated Calabi-Yau geometries arising here are more promising than the tori considered in simple models \cite{ADDpheno} because the absence of isometries allows KK modes to decay freely, without their scattering from branes.

\subsubsection*{Tests of Newton's inverse square law}

Precise tests of general relativity \cite{GRTests} provide another class of robust constraints on the models described here. These tests are sensitive only to states having masses in the sub-eV range or lower, but in some circumstances can be sensitive to interactions that are weaker than gravitational in strength.

Tests of Newton's inverse square law \cite{InvSqTests} over micron distances provide among the most robust tests. These are sensitive to two kinds of states: the large number of KK modes with sub-eV masses in large-hierarchy vacua \cite{ADDpheno}; and the various moduli that generically lie in this mass range for both large- and small-hierarchy models. Each of these can mediate a long-range force between test bodies over a range of order microns or larger.

The force mediated by exchange of moduli would deviate from an inverse square law and instead would follow the standard exponential Yukawa form for which experimenters search. Those moduli coupling with 4D gravitational strength, $\propto 1/M_p$, become constrained once the range of this force, $\sim 1/m$, becomes larger than $\sim 45$ $\mu$m \cite{InvSqTests}.

The signature expected from an exchange of a tower of KK modes in the large-hierarchy case has a slightly different signature, however, since the coherent sum over the KK tower never produces an exponential form; instead producing a crossover between the $1/r^2$ law at long distances to a $1/r^4$ law at short distances \cite{DevInvSqLaw}. The details of this crossover can depend on the precise shape of the two large dimensions, since they depend dominantly on the properties of the lightest KK modes.

\subsubsection*{Cosmology}

Cosmology of the very early Universe also furnishes strong constraints on any model with very large dimensions \cite{ADDpheno,SLEDpheno,MSLED}. This is because bulk KK modes of the large dimensions can ruin the success of Hot Big Bang cosmology if they (or their decay products) are too abundant at the epoch of Big Bang Nucleosynthesis (BBN) or thereafter. In particular, there is a `normalcy' temperature \cite{ADDpheno}, $T_\star$, above which the thermal history of the Universe on the SM brane is not simply described by the extrapolation of the standard Hot Big Bang to higher temperatures.

One way cosmology could become nonstandard is if the SM brane were to cool more quickly through evaporation into the bulk rather than expansion of the on-brane geometry. In the large-hierarchy case with $M_{6D} \sim 10$ TeV this occurs once $T \gsim 100$ MeV (and for the small-hierarchy case with $M_{10D} \sim 1$ TeV it instead happens once $T \gsim 10$ GeV) \cite{ADDpheno}. Since both of these are larger than BBN temperatures, $T_{\scriptscriptstyle BBN} \sim 1$ MeV, they need not be a problem. Rather, they are a complication when extrapolating to the earlier Universe (as is of interest, say, when finding a dark-matter candidate). In particular, not having a radiation-dominated thermal bath makes a WIMP description of dark matter less attractive.

Relics of bulk KK states are cosmologically dangerous, however. This is because the stabilisation of the extra dimensions imposes a comparatively large energy cost on changing the geometry's shape, making the KK modes lose energy with universal expansion like massive particles rather than radiation: $\rho \propto a^{-3}$. If too abundant, bulk states produced by thermal evaporation from a SM brane would carry too much energy and can over-close the Universe if $T$ is just a few MeV in the large-hierarchy case (with $M_{6D} \sim 10$ TeV), or $T \gsim 300$ MeV for small hierarchies (if $M_{10D} \sim 1$ TeV). Furthermore, if KK modes have an appreciable branching fraction into photons, even a decay rate as small as $T^3/M_p^2$ produces enough decays to be noticeable above backgrounds for temperatures in MeV ranges \cite{ADDpheno}.

These constraints largely require a pre-BBN history that suppresses the abundance of KK modes relative to those produced by thermal evaporation from the brane. It helps if they can decay invisibly once produced, as is required in any case from the neutron star bounds considered earlier.

A possible scenario might start with an inflationary epoch\footnote{If the overall volume were to evolve during inflation, see \cite{VolInf} for example, it could also happen that $M_s/M_p$ also evolves, allowing the string scale relevant to inflation to be much larger than the TeV scales appearing in the present Universe. } during which six dimensions grow, though with two of these ultimately stabilising at large dimensions while the visible four continue their growth into the present. Such an inflationary regime would iron out any wrinkles in the two large extra dimensions as well as from the four dimensions we see, effectively removing the dangerous KK modes from the initial conditions of the later Universe. It would also dilute otherwise the influence of the problematic moduli generically responsible for the Cosmological Modulus Problem \cite{CMP}. A viable cosmology could follow if this is followed by a reheat on the SM brane to temperatures not too far above nucleosynthesis temperatures, with the explanation for dark matter (and possibly baryons) potentially arising as relics from the reheating process. (See \cite{BaryDM} for a more detailed model in this spirit.)

Constructing such a scenario in detail would be worthwhile, but goes beyond the scope of the present paper. It is nevertheless encouraging that the basic ingredients likely to be required are present in these LARGE-volume constructions \cite{CM}.

\subsection{Less generic tests}

We next turn to potential signals that rely on the existence of particle states associated with the stringy UV completion, that are not generic consequences of supersymmetric large extra dimensions.

\subsubsection*{Moduli and precision tests of gravity}

The spectrum of moduli in these constructions generically involve several kinds of unusually light scalars. More remarkably, the masses of these scalars appear to be stable against radiative corrections. As described earlier, the mass of generic moduli lie in the $10^{-2}$ eV range that can be of interest to tests of Newton's inverse-square law over micron distances. But the mass of the volume and fibre moduli, $\delta \phi_2$ and $\delta\phi_1$, are much smaller, making them potentially relevant to tests of general relativity in the solar system and with binary pulsars.

Despite these small masses we believe these scalars are unlikely to be observed in present-day experiments. This is because the low masses for these states come hand-in-hand with the weakness of their coupling to the SM brane. (Recall for these purposes that it is the SM brane that breaks supersymmetry the most, and corrections to the scalar mass from SM particles go like $\zeta M_p/\vo$, where couplings to SM fields are of strength $\zeta/M_p$.) Since the discussion of \S3\ shows that $\zeta \lsim 1/\vo^{1/2} \sim 10^{-15}$ this means that the masses are only small enough to be of interest for terrestrial or solar system tests when the couplings are small enough to make the effects of scalar exchange too small to be measured. Conversely, should we have missed a graph that increases the couplings to $\zeta \sim O(1)$, the same kind of graph is likely also to lift the mass up to that of a generic modulus, $m \sim O(M_p/\vo)$.

\subsubsection*{Cosmology}

As we have seen, pre-BBN cosmology must be very different than a simple extrapolation of Hot Big Bang cosmology to higher temperatures. A proper identification of a dark-matter candidate requires a formulation of what this new cosmology is. The good news is that there are a number of candidate new particles that are not too heavy and do not couple appreciably to SM fields, and so could plausibly be a dark-matter candidate.

What is more surprising is that the lightest moduli can be light enough to be cosmologically active into the later Universe, and right into the present day for the large-hierarchy models with $m_1 \sim M_p/\vo^2 \sim 10^{-32}$ eV. This suggests that even the much-later Universe could be described by a scalar-tensor model, with its weaker-than-gravitational couplings explaining both the stability of its small mass against quantum corrections and the absence of evidence for new forces. The presence of such a light modulus resembles experience with six-dimensional models, for which the small scalar mass is related to the small size of the vacuum energy, leading to a late-time quintessence cosmology \cite{6Dcosmo} \footnote{For other interesting cosmological implications
of ultra-light scalars see \cite{Marsh}.}. It would be instructive to explore whether similar cosmologies are possible for the full string constructions, though a proper calculation requires a quantitative understanding of brane back-reaction.

\subsubsection*{Accelerator physics}

The detailed non-minimal spectrum predicted by the stringy UV completion also has implications for the signature of these kinds of models in collider experiments. In particular, there is more to be discovered than the minimal set that comes for any theory of supersymmetric large extra dimensions.

Among these other states are string states and those states associated with the KK scale of the small cycles and of the small base of the fibration, all of which lie at TeV scales. It is noteworthy that the 6D Planck mass, $M_{6D}$, is always much larger than either $M_s$ or $M_\KK^{10D}$ in these models. This is crucial for tests at colliders because it means that there can be observable signals, despite having the 6D Planck scale above 10 TeV. By contrast, for the pure graviton emission of simple ADD models, there are no observable signals if $M_{6D} > 10$ TeV.

A proper exploration of the precise signals to be seen requires a more detailed construction of the physics on the SM brane; a topic to be explored in further work.

\section{Conclusions}

Recent progress on modulus stabilisation has allowed explicit string theory realisations of the three main proposals to address the hierarchy problem: low-energy supersymmetry \cite{KKLT,LVS}, warped extra dimensions \cite{GKP} and large extra dimensions \cite{LVS}.

For large dimensions there is real added value to finding such a string embedding, complete with a modulus stabilising mechanism. This is because large extra dimensions in themselves do not solve the hierarchy problem \cite{ADD}; rather they move the problem to the problem of stabilising the extra dimensions at exponentially large values. By dynamically stabilising the extra dimensions, the string theory realisation completes the unfinished goal of solving the hierarchy problem.

The present work extends these constructions to include the most extreme and interesting case where two dimensions are on sub-micron scales \cite{ADD}. In this paper we find examples of Type IIB string vacua that stabilise the closed string moduli in a strongly anisotropic way, such that two dimensions are hierarchically larger than the others four. We identify two scenarios that achieve this, both using K3 or $T^4$-fibered Calabi-Yau compactifications.

In the first of these stabilisation is based on loop corrections to the K\"ahler potential, and the two large dimensions are only a few orders of magnitude larger than the rest. In the second, it is the presence of poly-instanton contributions to the superpotential that allow an exponential hierarchy to develop between the dimensions, that can be large enough to produce TeV string theory with micron-sized extra dimensions.

The need to choose fibred Calabi-Yau three-folds in order to realise our scenario relies on two properties of these compactification manifolds:
\begin{enumerate}
\item{}
K3 or $T^4$ fibrations are needed in order to have a hierarchically large two-cycle modulus $t_1$ keeping its dual four-cycle modulus $\tau_1$ small. For this to happen the volume $\mc{V}$, which is a cubic function of the two-cycle moduli $t_i$, has to depend on $t_1$ as $\mc{V}=t_1 f(t_j)+g(t_j)$ for $t_j\neq t_1$, since only in this case $\tau_1=\partial{\mc{V}}/\partial{t_1}$  will be independent of $t_1$ and can be kept small while having large $t_1$. This volume dependence on the 2-cycle moduli $t_i$ defines a K3 or $T^4$-fibred Calabi-Yau \cite{Schulz}.

\item{}
In our realisation of the large hierarchy scenario we needed a four-cycle which is not rigid in order not to have single-instanton contributions to the superpotential $W$ but only poly-instantons. This condition singles out K3 or $T^4$ surfaces since they admit Wilson lines and deformation moduli. An explicit realisation of poly-instantons in this set-up is left for future work.
\end{enumerate}

We provide a first look at the phenomenology of these models. Our preliminary exploration of their properties reveal that both types of scenarios predict a rich variety of observable phenomena on the edge of what can now be probed. Both types lead broadly to the phenomenology \cite{MSLED} of supersymmetric large extra dimensions \cite{SLED}, whose broad outlines consist of a very supersymmetric bulk ($m_{3/2} \sim 10^{-3}$ eV), weakly coupled to SM particles localised on a supersymmetry-breaking brane. Intriguingly, a robust prediction is the absence of MSSM superpartners, despite the presence of low-energy supersymmetry in the bulk.
Many of these features are generic to other TeV strings (which have been recently studied in detail in \cite{stst}).
But the modulus stabilising physics makes the phenomenology of our scenario much richer, because it provides a variety of other states that are not generic to SLED models, but which have interesting low-energy consequences. Among the differences from standard vanilla ADD scenario are:
\begin{itemize}
\item{}
The bulk is supersymmetric with supersymmetry broken at sub-eV scales. Supersymmetry is broken at the TeV scale at the standard model brane, in particular there are no supersymmetric partners of the standard model particles. This is also a  property of the SLED scenario \cite{SLED} but not in the original large extra dimension scenario \cite{ADD}. The next three properties are not present in the SLED scenario though.

\item{}
There are a variety of Kaluza-Klein and string states, all close to the TeV scale, whose masses are below the 6D Planck mass (whose value is normally taken as the benchmark for detection of extra-dimensional models at colliders).

\item{}
There is a rich spectrum of very light moduli with unusually small masses and couplings weaker than gravitational strength, such that they are consistent with present observations but with potential cosmological and astrophysical implications.
\end{itemize}

Because all of the different approaches to the hierarchy problem are realised in Type IIB string theory, it is tempting to seek cases where the different mechanisms address different hierarchies.

An important complication in the strongly anisotropic case is the presence of brane back-reaction, which for codimension-2 branes compete with standard mechanisms at a level that seriously complicates understanding the low-energy vacuum dynamics. If the new dynamics should provide a mechanism for understanding the small present-day vacuum energy, the presence of this new dynamics may be a blessing in disguise. Perhaps it is not a coincidence that back-reaction is strongest in the very anisotropic regime, which is also the case for which $M_\KK^{6D}$ is of order the observed dark energy scale. These special features of two large dimensions underlies the supersymmetric large extra dimensions scenario (SLED) \cite{SLED}, which generally requires not only exponentially large extra dimensions but precisely two exponentially large dimensions, as we find here.

We finally point out that we estimated the closed-string loop corrections to the scalar potential
to scale as $\vo^{-4}\sim 10^{-120} M_p^4 \sim \Lambda_{cc}^4$, reproducing
the contribution of loops of bulk states that in the standard 6D SLED scenarios give rise to the observed value of the cosmological constant.
However our stringy embedding differs from the standard 6D SLED scenarios since the vacuum energy is dominated by $\alpha'$ and non-perturbative effects
which give rise to an AdS minimum at order $\vo^{-3}$.
If the lifting mechanism cancels this order $\vo^{-3}$ contribution to the scalar potential,
the other corrections are smaller than $\vo^{-4}$.
This is clearly not enough to address the dark energy problem but together with the cancelation mechanisms of \cite{SLED} it may give rise to a stringy scenario for dark energy.

\bigskip

{\noindent}{\bf Acknowledgements}

We would like to thank Massimo Bianchi, Marco Bill\'o, Ralph Blumenhagen, Andres Collinucci, Joe Conlon, Mark Goodsell, Arthur Hebecker,
Sven Krippendorf, Anshuman Maharana, Fernando Marchesano, Luca Martucci, Francisco Pedro, Roberto Valandro and Timo Weigand for useful discussions.
We thank the Abdus Salam International Centre for Theoretical Physics and Perimeter Institute for Theoretical Physics for providing resources that several times allowed us to meet at one place and devote relatively undivided time to this project. CB's research is supported in part by funds from the Natural Sciences and Engineering Research Council (NSERC) of Canada. Research at the Perimeter Institute is supported in part by the Government of Canada through NSERC and by the Province of Ontario through MEDT.

\end{document}